\documentclass[aps,prd,preprint,superscriptaddress,nofootinbib,longbibliography]{revtex4-1}
\usepackage{epsfig,dsfont,amssymb,amsmath,amsthm,amsfonts,amsbsy,mathrsfs,mathtools}
\usepackage{graphicx}

\usepackage{float}

\usepackage[normalem]{ulem}
\usepackage{color}
\begin{document}

\title{Scalar perturbations on the background of Kerr black holes  in the quadratic dynamical Chern-Simons gravity}

\author{Yuan-Xing Gao}
%\email{158059238@qq.com}
\author{Yang Huang}
\author{Dao-Jun Liu}
\email{djliu@shnu.edu.cn}
\affiliation{Center for Astrophysics and Department of Physics, Shanghai Normal University, 100 Guilin Road, Shanghai 200234, China}
\begin{abstract}
	We study the scalar perturbation on the background of a Kerr black hole in the dynamical Chern-Simons modified gravity with a quadratic coupling between the scalar field and Chern-Simons term.  In particular,  the late-time tails of scalar perturbations are investigated numerically in time domain  by using the hyperboloidal foliation method.  It is found that the Kerr black hole becomes unstable under linear perturbations in a certain region of the parameter space, which depends on the  harmonic azimuthal index $m$ of the perturbation's mode. This may indicate  that some Kerr black holes in this theory will get spontaneously scalarized into a non-Kerr black hole.
\end{abstract}
\maketitle

\section{introduction}

Einstein's theory of general relativity (GR) has passed all precision tests performed so far with flying colors \cite{Will2014}. However, due to its incompatibility with quantum theory and motivations from cosmology, it is believed that GR may not be the final theory to describe the  gravitational physics but just an effective description of an unknown fundamental theory of gravity and should be modified at both low and high energies \cite{0264-9381-32-24-243001}.  To go beyond GR, plenty of alternative theories of gravity have been proposed, see, e.g., Ref. \cite{Faraoni2011} for a comprehensive review. 

Among various modifications or extensions of GR, a class of theories, the so-called quadratic gravity of which the action contains terms quadratic in the curvature, is of particular interest. It is known that a major obstacle on the road to quantum gravity is that GR cannot be perturbatively renormalized as the standard model of particle physics. The situation changes if the Einstein-Hilbert action is assumed to be only the first term in an expansion containing high-order curvature invariants. In fact, Stelle showed a long time ago that including quadratic curvature terms in the action makes the theory renormalizable \cite{PhysRevD.16.953}. These extra  quadratic curvature terms lead to new effects in the strong-field regime, manifesting themselves most naturally in the black hole (BH) solutions of these models.

Chern-Simons (CS) modified gravity \cite{PhysRevD.68.104012} is a special kind of quadratic gravity in which an additonal CS invariant  (i.e., the contraction of
the Riemann tensor and its dual, also called Pontryagin density) coupled to a scalar field is added in the action, which captures leading-order, gravitational parity violation.\footnote{The CS invariant term plays an interesting role in gravity even in the case in which its coupling is a constant. Indeed, in that case, it modifies the surface terms of the theory with consequences at the level of the holographic description of the system \cite{Miskovic:2009bm}.} Such a term is not only reduced from the Green-Schwarz anomaly canceling mechanism in heterotic string theory \cite{CAMPBELL1991778,Moura2006} but also appears naturally in loop quantum gravity \cite{Ashtekar:1988sw}, especially when the Barbero-Immirzi parameter is promoted to a scalar field coupled to the Nieh-Yan invariant \cite{Mercuri:2007ki,Taveras:2008yf,Calcagni:2009xz}. 
For a review of CS modified gravity, we refer to Ref. \cite{ALEXANDER20091}.

The CS modified gravity was at first investigated in the nondynamical formulation, in which  there is no kinetic term for the scalar field in the action, and hence it is assumed  to be an \emph{a priori} prescribed spacetime function. However, nondynamical CS theory is quite contrived because a valid solution for the spacetime must satisfy the condition that the Pontryagin
density vanishes.  It has been shown from different aspects that nondynamical CS modified gravity is theoretically problematic \cite{PhysRevD.79.084043,Grumiller:2007rv,Yunes:2007ss}. Therefore, it can only be considered as a toy model used to gain some insight in parity-violating theories of gravity.

In the last decade, much attention has been payed to the so-called dynamical Chern-Simons (dCS) modified gravity \cite{Smith:2007jm}, which is a more natural formulation, in which the scalar field is treated as a dynamical field. It is worth noting that although the action of the nondynamical CS gravity  can be obtained as a certain limit of that of dCS gravity, the nondynamical CS gravity and dCS gravity are inequivalent and independent theories. 

When the spacetime has spherical symmetry, the parity-violating CS invariant vanishes, and then the Schwarzschild solution with vanishing scalar field is an exact solution of dCS gravity. The perturbations of Schwarzschild BHs in dCS modified gravity were first investigated by Cardoso and Gualtier \cite{PhysRevD.80.064008}. 
Later, Garfinkle \emph{et al}. found that dCS modified gravity is linearly stable on Schwarzschild and other of physically relevant backgrounds by performing a linear stability analysis in the geometric optics approximation and discussed the speed of gravitational waves in this theory \cite{PhysRevD.82.041501}. The linear mode stability for a generic massive scalar in the background of a Schwarzschild BH in dCS gravity was proved recently \cite{Kimura:2018nxk}.

The rotating BH solutions in dCS gravity have been obtained in the small-coupling and/or slow-rotation limit  by many authors \cite{PhysRevD.79.084043,Konno2009,PhysRevD.84.124033,PhysRevD.90.044061,Yagi:2012ya,McNees2016}. The null geodesics and shadow of a slowly rotating BH in dCS
gravity with a small coupling constant are also studied \cite{Amarilla2010}.  Chen and Jing investigated the geodetic precession and the strong gravitational lensing in the slowly rotating BH in the dCS gravity and found the effects of the CS coupling parameter on the geodetic precession angle for the timelike particles and the coefficients of gravitational lensing in the strong-field limit \cite{Chen2010}. The perturbative BH solutions have the advantage
of having analytic expressions, leading to some insights on the effect of the CS coupling.
However, some important features, occurring in the fast spinning and/or large coupling regimes, cannot be captured by them. Recently, spinning BHs in dCS gravity were constructed by directly solving the field equations, without resorting to any perturbative expansion \cite{Delsate:2018ome}.

The possible signatures of dCS gravity in the gravitational-wave emission produced in the inspiral of stellar compact objects into massive BHs are investigated in Ref.\cite{Sopuerta2009}, for both intermediate- and extreme-mass ratios. By applying the  effective field theory (EFT) method, Loutrel \emph{et al}. \cite{Loutrel:2018ydv} recently derived the leading post-Newtonian order spin-precession equations for binary BHs in dCS gravity. It is worth mentioning that  the detection of gravitational waves has ruled out a lot of alternative theories\cite{Baker2017,Creminelli2017,Sakstein:2017xjx,PhysRevLett.119.251304} (see Refs \cite{Lombriser2016,Lombriser2017} for some earlier works), however,  as is mentioned in Ref. \cite{PhysRevD.97.084005}, gravitational waves in dCS gravity propagate at the speed of light on conformally flat background spacetimes, such as Friedmann-Robertson-Walker spacetime  \cite{PhysRevD.82.041501,0264-9381-33-5-054001}. 

Hitherto, most of the literature on dCS modified gravity considers the linear coupling between the scalar field and the CS term, because of the parity and other symmetry reasons.  It is known that a linear coupling means the field respects a shift symmetry and the theory is protected from acquiring a mass for the field.
In this paper, inspired by the phenomenon of spontaneous scalarization
recently discussed in the so-called quadratic scalar--Gauss-Bonnet gravity \cite{Antoniou2018,PhysRevLett.120.131103,PhysRevLett.120.131104,Antoniou:2017hxj}, we want to consider a dCS modified theory of gravity in which the CS invariant is coupled to the quadratic function of the dynamical scalar field.   What we are concerned about is the stability of the Kerr black hole in this theory and  how a perturbation of the scalar field coupled with the CS invariant evolves in the Kerr background.

The paper is organized as follows. In Sec.\ref{sec:2}, a brief review on the dCS theory of gravity with a quadratic coupling is given. Next, in Sec.\ref{sec:3}, we derive the equation of motion for the scalar perturbations. The numerical method employed is described in Sec.\ref{sec:4} and the main results are presented in Sec.\ref{sec:result}. An analysis of the validity of taking the theory as an EFT  is performed in Sec.\ref{EFT}. Finally,  we conclude in Sec.\ref{sec:conclusion}. Unless otherwise stated, we use geometric units with $G=c=1$ and the mostly plus metric signature.

\section{dynamical Chern-Simons gravity with  a quadratic coupling}
\label{sec:2}
A general model of dCS modified gravity can be described by the action
\begin{equation}\label{eq:action}
\begin{aligned}
S=\int d^4x\sqrt{-g}\left[\kappa R-\frac{1}{2}\nabla_\mu\Phi\nabla^\mu\Phi-V(\Phi)+\tilde{\alpha} f(\Phi)^*RR +\mathcal{L}_m \right]
\end{aligned}
\end{equation}
where $\kappa=(16\pi )^{-1}$, $\Phi$ is a real scalar field with a self-interaction potential $V(\Phi)$,  $\tilde{\alpha}$ is a coupling constant with dimension of $(length)^2$, $f(\Phi)$ is an arbitrary dimensionless function of the scalar field that itself is a dimensionless quantity, and $\mathcal{L}_m$ denotes the Lagrangian for matter that minimally coupled to gravity.  As usual, $g$ denotes the determinant of the metric $g_{\mu\nu}$, and $R$ is
the Ricci scalar. The CS invariant is defined by
\begin{equation}
^*RR=\frac{1}{2}\epsilon^{\alpha\beta\gamma\delta}{R^\mu}_{ \nu\gamma\delta}{R^\nu}_{\mu\alpha\beta},
\end{equation} 
where $\epsilon^{\alpha\beta\gamma\delta}$ and ${R^\mu}_{ \nu\gamma\delta}$ are the four-dimensional Levi-Civit\`{a} tensor and Riemann curvature tensor, respectively. Note that the CS invariant itself is a topological term and can be expressed as a total divergence \cite{ALEXANDER20091}.  In this work, we omit the contribution of matter and only consider the vacuum solution. Moreover, for simplicity, the scalar field is assumed to be massless and has no self-interaction, i.e., $V(\Phi)=0$.  

The equation of motion for the metric $g_{\mu\nu}$  derived from the action with vanishing $V(\Phi)$ and $\mathcal{L}_m$ in Eq.\eqref{eq:action} are the modified Einstein equation  
\begin{equation}\label{eq:Einstein}
G^{\mu\nu}+4\frac{\tilde{\alpha}}{\kappa}C^{\mu\nu}=\frac{1}{2\kappa}T^{\mu\nu}_{(\Phi)}
\end{equation}
where $G^{\mu\nu}$ is the contravariant Einstein tensor. The tensor $C^{\mu\nu}$ and the stress-energy
tensor for the scalar field $T^{\mu\nu}_{(\Phi)}$ are defined by
\begin{equation}\label{eq:C-tensor}
C^{\mu\nu}=\nabla_\sigma f(\Phi)\epsilon^{\sigma\alpha\beta(\mu}\nabla_\beta {R^{\nu)}}_\alpha+\nabla_\alpha\nabla_\beta f(\Phi)^*R^{\alpha(\mu\nu)\beta}
\end{equation}
and
\begin{equation}
T^{\mu\nu}_{(\Phi)}=\nabla^\mu\Phi\nabla^\nu\Phi-\frac{1}{2}g^{\mu\nu}(\nabla_\lambda\Phi)(\nabla^\lambda\Phi),
\end{equation}
respectively. 
On the other hand, the Klein-Gordon equation for the scalar field $\Phi$ is modified to be
\begin{equation}\label{eq:KG}
\Box\Phi+\tilde{\alpha}\,^*RR\frac{\mathrm{d}f(\Phi)}{\mathrm{d}\Phi}=0.
\end{equation}
By taking the covariant divergence of Eq.\eqref{eq:C-tensor}, it is not difficult to find that 
%Considering the consistency condition, we take the covariant divergence of Eq.$(4)$
\begin{equation}
\nabla_\mu C^{\mu\nu}=\frac{^*RR}{8}\nabla^{\nu}f(\Phi).
\end{equation}
Therefore, from Eq.\eqref{eq:Einstein}, the evolution of $\Phi$  is also determined by 
\begin{equation}\label{eq:e-mConserv}
\tilde{\alpha}\,^*RR\nabla^\nu f(\Phi)=\nabla_\mu T^{\mu\nu}_{(\Phi)},
\end{equation}
which is just the requirement of energy-momentum conservation and equivalent to Eq.\eqref{eq:KG}.  

As is mentioned in the Introduction, in the present paper,  we simply choose
\begin{equation}\label{eq:f}
f(\Phi)=\Phi^2.
\end{equation} 
We shall call such a theory quadratic dCS gravity.  Clearly, it exhibits  $\Phi\to -\Phi$ symmetry, which is of interest in the field theory context. Moreover, in quadratic dCS gravity, the field equation for $\Phi$ is a linear homogeneous differential equation, so it will be very convenient to study the perturbation in the decoupling limit.

On the other hand, it is not difficult to find that for the case $f(\Phi)=\Phi^2$ Eq.\eqref{eq:Einstein} reduces to $G^{\mu\nu}=0$ when $\Phi=\Phi_0$ is a constant, which means all the vacuum solutions in GR, including the Kerr solution, will be recovered. Unfortunately, $\Phi=\Phi_0$ is not a solution of Eq.\eqref{eq:KG} unless $\Phi_0=0$. 

\section{equation of motion for scalar perturbations on the Kerr background}
\label{sec:3}
Among the vacuum solutions in GR, the Kerr solution, which describes a stationary, axisymmetric spacetime, such as that around a Kerr black hole, is of most interest. In Boyer-Lindquist coordinates, the line element of Kerr spacetime reads 
\begin{equation}
\begin{aligned}
ds^2=-\frac{\Delta}{\rho^2}(dt-a\sin^2\theta d\phi)^2+\frac{\sin^2\theta}{\rho^2}[(r^2+a^2)d\phi-adt]^2
+\frac{\rho^2}{\Delta}dr^2+\rho^2d\theta^2,
\end{aligned}
\end{equation}
where
\begin{displaymath}
\begin{aligned}
\Delta&\equiv r^2-2Mr+a^2,\\
\rho^2&\equiv r^2+a^2\cos^2\theta,\\
a&\equiv\frac{J}{M},
\end{aligned}
\end{displaymath}
and $M$ and $J$ are the Arnowitt-Deser-Misner (ADM) mass and ADM angular momentum, respectively.

We shall investigate  whether there is a regime within which the Kerr black hole solution is unstable in the framework of the quadratic dCS gravity. To this end, it is useful to investigate the linear perturbations of the Kerr spacetime with the trivial scalar field. Fortunately,  similar to the theory discussed in Ref.\cite{PhysRevLett.120.131103},  the equations governing the perturbations of the metric $\delta g_{\mu\nu}$ are decoupled from that of the scalar field in the linear level. Therefore, we focus on the perturbation of the scalar field $\delta\Phi$, which is governed by the following equation
\begin{equation}\label{eq:perturbedKG}
(\Box+2\tilde{\alpha}^*RR)\delta\Phi=0.
\end{equation} 
This is a Teukolsky-like equation, and it is obvious that the curvature correction acts as an effective mass. For the Kerr spacetime, the CS invariant reads
\begin{equation}
^*RR=\frac{96aM^2r\cos\theta(3r^2-a^2\cos^2\theta)(r^2-3a^2\cos^2\theta)}{(r^2+a^2\cos^2\theta)^6}.
\end{equation}
Since its sign could be positive or negative, the curvature correction term may provide a negative effective mass squared which may cause a tachyon  instability in certain regime. 

To solve Eq.\eqref{eq:perturbedKG} numerically, it is helpful to rewrite the Kerr metric in the ingoing Kerr-Schild coordinates $\{\tilde{t},r,\theta,\varphi\}$ through the following transformation:
\begin{equation}
\begin{aligned}
&\mathrm{d}\tilde{t}=\mathrm{d}t+\frac{2Mr}{\Delta}\mathrm{d}r,\\
&\mathrm{d}\varphi=\mathrm{d}\phi+\frac{a}{\Delta}\mathrm{d}r.
\end{aligned}
\end{equation}
As a result, the line element of Kerr metric and the equation for scalar perturbation \eqref{eq:perturbedKG} can be rewritten as
\begin{equation}
\begin{aligned}
\mathrm{d}s^2=&-\left(1-\frac{2Mr}{\rho^2}\right)\mathrm{d}\tilde{t}^2-\frac{4aMr}{\rho^2}\sin^2\theta \mathrm{d}\tilde{t}\mathrm{d}\varphi+\frac{4Mr}{\rho^2}\mathrm{d}\tilde{t}\mathrm{d}r\\
&+\left(1+\frac{2Mr}{\rho^2}\right)\mathrm{d}r^2-2a\sin^2\theta\left(1+\frac{2Mr}{\rho^2}\right)\mathrm{d}r\mathrm{d}\varphi+\rho^2\mathrm{d}\theta^2\\
&+\left(r^2+a^2+\frac{2Ma^2r\sin^2\theta}{\rho^2}\right)\sin^2\theta \mathrm{d}\varphi^2
\end{aligned}
\end{equation}
and 
\begin{equation}\label{eq:KG2}
\begin{aligned}
(\rho^2+2Mr)\partial_{\tilde{t}}^2\delta\Phi=&2M\partial_{\tilde{t}}\delta\Phi+4Mr\partial_{\tilde{t}}\partial_r\delta\Phi+\partial_r(\Delta\partial_r\delta\Phi)\\
&+2a\partial_r\partial_\varphi\delta\Phi+\frac{1}{\sin^2\theta}\partial_\varphi^2\delta\Phi+\frac{1}{\sin\theta}\partial_\theta(\sin\theta\partial_\theta\delta\Phi)+2\tilde{\alpha}\rho^2\;{^*R}R\delta\Phi,
\end{aligned}
\end{equation}
respectively. Given the axial symmetry of the Kerr geometry, the perturbative variable $\delta\Phi$ can be decomposed as 
\begin{equation}
\delta\Phi(\tilde{t},r,\theta,\varphi)=\frac{1}{r}\psi(\tilde{t},r,\theta) e^{im\varphi}.
\end{equation}
 Inserting the above expression into Eq.\eqref{eq:KG2}, we finally obtain that
\begin{equation}\label{eq:TE21}
\begin{aligned}
&A^{\tilde{t}\tilde{t}}\partial_{\tilde{t}}^2\psi+A^{\tilde{t}r}\partial_{\tilde{t}}\partial_r\psi+A^{rr}\partial^2_r\psi+A^{\theta\theta}\partial_\theta^2\psi+B^{\tilde{t}}\partial_{\tilde{t}}\psi
+B^r\partial_r\psi+B^\theta\partial_\theta\psi+C\psi=0,
\end{aligned}
\end{equation}
where
\begin{equation}
\begin{aligned}
&A^{\tilde{t}\tilde{t}}=\rho^2+2Mr,\\
&A^{\tilde{t}r}=-4Mr,\\
&A^{rr}=-\Delta,\\
&A^{\theta\theta}=-1,\\
&B^{\tilde{t}}=2M,\\
&B^r=\frac{2}{r}(a^2-Mr)-2ima,\\
&B^\theta=-\cot\theta,\\
&C=\frac{m^2}{\sin^2\theta}-\frac{2(a^2-Mr)}{r^2}-2\tilde{\alpha}\rho^2{^*RR}+\frac{2ima}{r}.
\end{aligned}
\end{equation}

\section{numerical method}
\label{sec:4}

Equation\eqref{eq:TE21} is a modified homogeneous 2+1 Teukolsky equation for spin-$0$ perturbations. 
Solving this partial differential equation (PDE) is not a trivial task. In fact, it was not long ago that the behavior of a scalar field on fixed Kerr background in GR was examined by R\'{a}cz and T\'{o}th \cite{Racz2011}. In their work, a numerical framework incorporating the techniques of conformal compactification and hyperbolic initial value formulation is employed. Later, a numerical solution of the 2 + 1 Teukolsky equation for generic spin perturbations on a hyperboloidal and horizon penetrating foliation of Kerr was also investigated \cite{0264-9381-30-11-115013,0264-9381-31-24-245004,ZENGINOGLU20112286}.  

In the present work, we shall  solve the Eq.\eqref{eq:TE21} by using the hyperboloidal foliation method. For this purpose, we first introduce the compactified radial coordinate $R$ and the suitable time coordinate $T$ with the following definitions following R\'{a}cz and T\'{o}th \cite{Racz2011}
\begin{equation}
\begin{aligned}
\tilde{t}=T+h(R),\ \ r=\frac{R}{\Omega(R)},
\end{aligned}
\end{equation}
where
\begin{equation}
h(R)=\frac{1+R^2}{2\Omega}-4M\ln(2\Omega)
\end{equation}
and
\begin{equation}
\Omega(R)=\frac{1-R^2}{2}.
\end{equation}
The event horizon $R_+$ in the new radial coordinate $R$ is located at
\begin{equation}
R_+=\frac{2\sqrt{2M\sqrt{M^2-a^2}-a^2+2M^2+1}-2}{2(\sqrt{M^2-a^2}+M)}.
\end{equation}
In addition, we can further define the boost function $H(R)$, which is useful for later computation:
\begin{equation}
H=\frac{dh}{dr}(R).
\end{equation}
Then, putting  these relations into the Teukolsky-like equation \eqref{eq:TE21} and bearing in mind that
\begin{equation}
\partial_{\tilde{t}}=\partial_T,\ \ \partial_{r}=-H\partial_T+\frac{2\Omega^2}{1+R^2}\partial_R,
\end{equation}
we finally obtain that 
%\begin{equation}\label{eq:TE22}
%\begin{aligned}
%&A^{TT}\partial^2_T\psi+A^{TR}\partial_T\partial_R\psi+A^{RR}\partial^2_R\psi+A^{\theta\theta}\partial^2_\theta\psi\\
%&+B^T\partial_T\psi+B^R\partial_R\psi+B^\theta\partial_\theta\psi+C\psi=0,
%\end{aligned}
%\end{equation}
\begin{equation}\label{eq:TE23}
\begin{aligned}
&\partial^2_T\psi+\tilde{A}^{TR}\partial_T\partial_R\psi+\tilde{A}^{RR}\partial^2_R\psi+\tilde{A}^{\theta\theta}\partial^2_\theta\psi+\tilde{B}^T\partial_T\psi\\
&+\tilde{B}^R\partial_R\psi+\tilde{B}^\theta\partial_\theta\psi+\tilde{C}\psi=0\\
\end{aligned}
\end{equation}
where
\begin{align}
\{\tilde{A}^{TR},\tilde{A}^{RR},\tilde{A}^{\theta\theta},\tilde{B}^{T},\tilde{B}^{R},\tilde{B}^{\theta},\tilde{C}\}=\frac{1}{A^{TT}}\{A^{TR},A^{RR},A^{\theta\theta},B^{T},B^{R},B^{\theta},C\}
\end{align}
and
\begin{align}\label{eq:26}
&A^{TT}=A^{\tilde{t}\tilde{t}}-HA^{\tilde{t}r}+H^2A^{rr},\\
&A^{TR}=\frac{2\Omega^2}{1+R^2}A^{\tilde{t}r}-\frac{4\Omega^2}{1+R^2}HA^{rr},\\
&A^{RR}=\left(\frac{2\Omega^2}{1+R^2}\right)^2A^{rr},\\
&B^T=B^{\tilde{t}}-HB^{r}-H'\left(\frac{2\Omega^2}{1+R^2}\right)A^{rr},\\
&B^R=\frac{2\Omega^2}{1+R^2}\left [B^{r}+\left(\frac{2\Omega^2}{1+R^2}\right)'A^{rr}\right],
\end{align}
where the prime denotes the derivative with respect to $R$. 

It is convenient to  introduce an auxiliary field $\Pi$ \cite{PhysRevD.56.3395} 
\begin{equation}
\Pi\equiv\partial_T\psi+b\partial_R\psi,\\
\end{equation}
where the coefficient 
\begin{equation}
b\equiv\frac{\tilde{A}^{TR}+\sqrt{(\tilde{A}^{TR})^2-4\tilde{A}^{RR}}}{2}.
\end{equation}
Then, Eq.\eqref{eq:TE23} can be converted into a coupled set of first-order equations in space and time,
\begin{equation}\label{eq:TE24}
\partial_T\left [{\begin{array}{*{20}c}
	\psi\\
	\Pi \\
	\end{array}} \right] + \left [{\begin{array}{*{20}c}
	\alpha_{11} & \alpha_{12} \\
	\alpha_{21} & \alpha_{22} \\
	\end{array}} \right] 
\partial_R\left [{\begin{array}{*{20}c}
	\psi\\
	\Pi \\
	\end{array}} \right]+
\left [{\begin{array}{*{20}c}
	\beta_{11} & \beta_{12} \\
	\beta_{21} & \beta_{22} \\
	\end{array}} \right]\left [{\begin{array}{*{20}c}
	\psi\\
	\Pi \\
	\end{array}} \right]=0
\end{equation}
where
\begin{equation}
\begin{matrix*}[l]
\alpha_{11}=b, & \beta_{11}=0,\\
\alpha_{12}=0, & \beta_{12}=-1,\\
\alpha_{21}=\tilde{B}^R+(b-\tilde{A}^{TR})\partial_R b-b\tilde{B}^T, &\beta_{21}=\tilde{A}^{\theta\theta}\partial^2_\theta+\tilde{B}^\theta\partial_\theta+\tilde{C}, \\
\alpha_{22}=\tilde{A}^{TR}-b, & \beta_{22}=\tilde{B}^\tau. 
\end{matrix*}
\end{equation}

When $\psi$ and $\Pi$ are split into real and imaginary parts as 
\begin{equation}
\psi=\psi_\mathcal{R}+i\psi_\mathcal{I}
\end{equation}
and
\begin{equation}
\Pi =\Pi_\mathcal{R}+i\Pi_\mathcal{I}, 
\end{equation}
Eq.\eqref{eq:TE24} can be written in the matrix form
\begin{equation}\label{eq:TE25}
\frac{du}{dT}=-(G\partial_R +Y+X)u
\end{equation}
where $u\equiv [ \psi_\mathcal{R}, \psi_\mathcal{I}, \Pi_\mathcal{R}, \Pi_\mathcal{I}]^{-1}$,
\begin{equation}
G\equiv\left [{\begin{array}{*{20}c}
	b & 0 & 0 & 0\\
	0 & b & 0 & 0\\
	\alpha_{21}^\mathcal{R} & -\alpha_{21}^\mathcal{I} & \alpha_{22} & 0\\
	\alpha_{21}^\mathcal{I} & \alpha_{21}^\mathcal{R} & 0 & \alpha_{22}
	\end{array}} \right], 
\end{equation}
\begin{equation}
Y\equiv\left [{\begin{array}{*{20}c}
	0 & 0 & 0 & 0\\
	0 & 0 & 0 & 0\\
	l_{31} & 0 & 0 & 0\\
	0 & l_{31} & 0 & 0
	\end{array}} \right], 
\end{equation}
and
\begin{equation}
X\equiv\left [{\begin{array}{*{20}c}
	0 & 0 & -1 & 0\\
	0 & 0 & 0 & -1\\
	\beta_{21}^\mathcal{R} & -\beta_{21}^\mathcal{I} & \beta_{22}^\mathcal{R} & -\beta_{22}^\mathcal{I}\\
	\beta_{21}^\mathcal{I} & \beta_{21}^\mathcal{R} & \beta_{22}^\mathcal{I} & \beta_{22}^\mathcal{R}
	\end{array}} \right].
\end{equation}
Note that all the coefficients are functions of coordinates $R$ and $\theta$.

Once the dynamical equation \eqref{eq:TE25} for scalar perturbations is derived,  the next task is to solve it numerically by employing a standard fourth-order Runge-Kutta integrator. Taking into account the computational efficiency, we compute the equation in a domain $(R_+,\ 1)\times(0,\ \pi)$ with grids of $801\times67$ points.
Before we conduct the numerical computation, it is necessary to define the new dimensionless coupling constant $\alpha$ by
\begin{equation}
\alpha=\tilde{\alpha}/M^2
\end{equation}
where $M$ is the BH's mass. In the actual process of computation, we set $M = 1$ for convenience.
%\subsection{Spatial derivitives}
To discretize the spatial parts, we use the fourth-order accurate finite differences formula \cite{PhysRevD.72.084022} in both the radial and angular directions.
%\subsection{Initial data}
We choose  spherically harmonic Gaussian bells centered at $R_c$ in the $R$ direction with different degrees of concentration  as the initial perturbations.  That is, the function $\psi$ initially takes the form
\begin{equation}\label{eq:IC}
\psi(t=0,R,\theta)\sim Y_{lm}\exp\left[\frac{-(R-R_c)^2}{2\sigma^2}\right]
\end{equation}
where $Y_{lm}$ represents the $\theta$-dependent spherical harmonics. Because of the relations presented  in Eqs.\eqref{eq:26}, the initial form of $\Pi$ is determined by
\begin{equation}
\Pi(t=0,R,\theta)=b\;\partial_R\psi(0,R,\theta).
\end{equation}

%\subsection{Boundary condition}
As for the boundary conditions, in the $R$ direction, the coefficients of the PDEs are always regular, and the use of hyperboloidal foliation and compactification achieves that the transformed system is pure outgoing at the outer boundary, which is just what we are looking for, so we do not have to exert the boundary conditions at infinity by hand anymore.
Similarly, since the foliation is horizon penetrating, we do not need to specify the inner boundary condition on the horizon, either \cite{ZENGINOGLU20112286}.

However, in the $\theta$ direction, the coefficients of the PDEs become singular at the pole where $\theta=0$ and $\pi$, and we need to use a staggered grid to avoid the inherent difficulties of evaluating expressions in which $\csc\theta$ is present \cite{PhysRevD.72.084022}. In the staggered grid, the values for $\theta=0$ and $\theta=\pi$ are located between two grid points. The points to the left (right) of $\theta=0\; (\theta=\pi)$ are considered ghost points, because these points are only used to impose the boundary conditions. In our fourth-order Runge-Kutta integrator, four ghost points are needed, two of which are in the front of the points where $\theta=0$ and the other two of which are at the back of the points where $\theta=\pi$. The values in the ghost points are updated according to the following strategies:
\begin{equation}
\left.
\begin{aligned}
\psi(T,R,\theta)&=\psi(T,R,-\theta)\\
\psi(T,R,\pi+\theta)&=\psi(T,R,\pi-\theta)
\end{aligned}
\right\}\;\mathrm{for}\;\ m=0,\pm2,\cdots
\end{equation}
and
\begin{equation}
\left.
\begin{aligned}
\psi(T,R,\theta)&=-\psi(T,R,-\theta)\\
\psi(T,R,\pi+\theta)&=-\psi(T,R,\pi-\theta)
\end{aligned}
\right\}\;\mathrm{for}\;\ m=\pm1,\pm3,\cdots
\end{equation}
%\subsection{The error of methods adopted in pure Kerr background}

\section{results}
\label{sec:result}

\begin{figure}        
	\centering 
	\includegraphics[width=0.48\textwidth]  {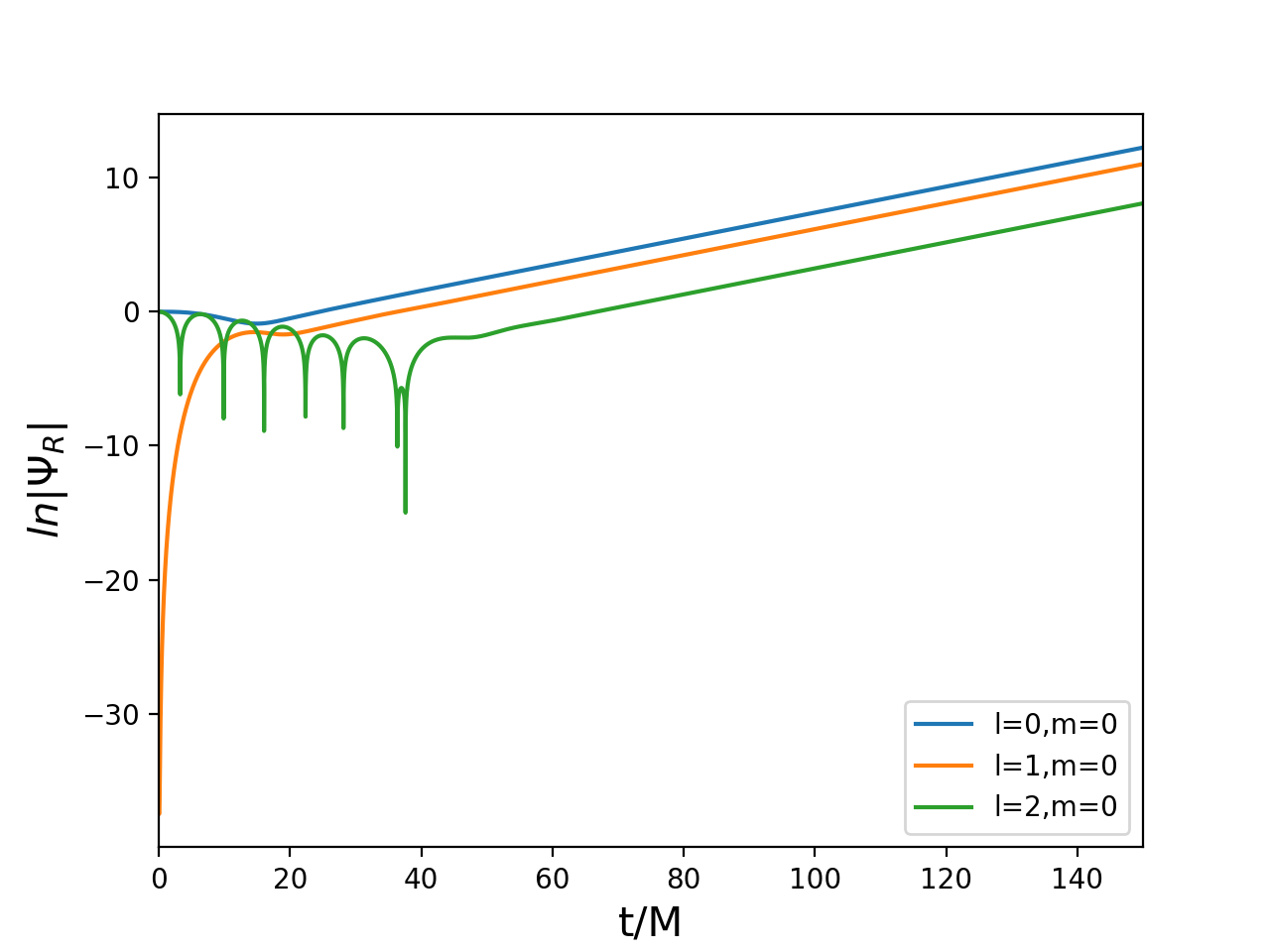}  
	\includegraphics[width=0.48\textwidth]  {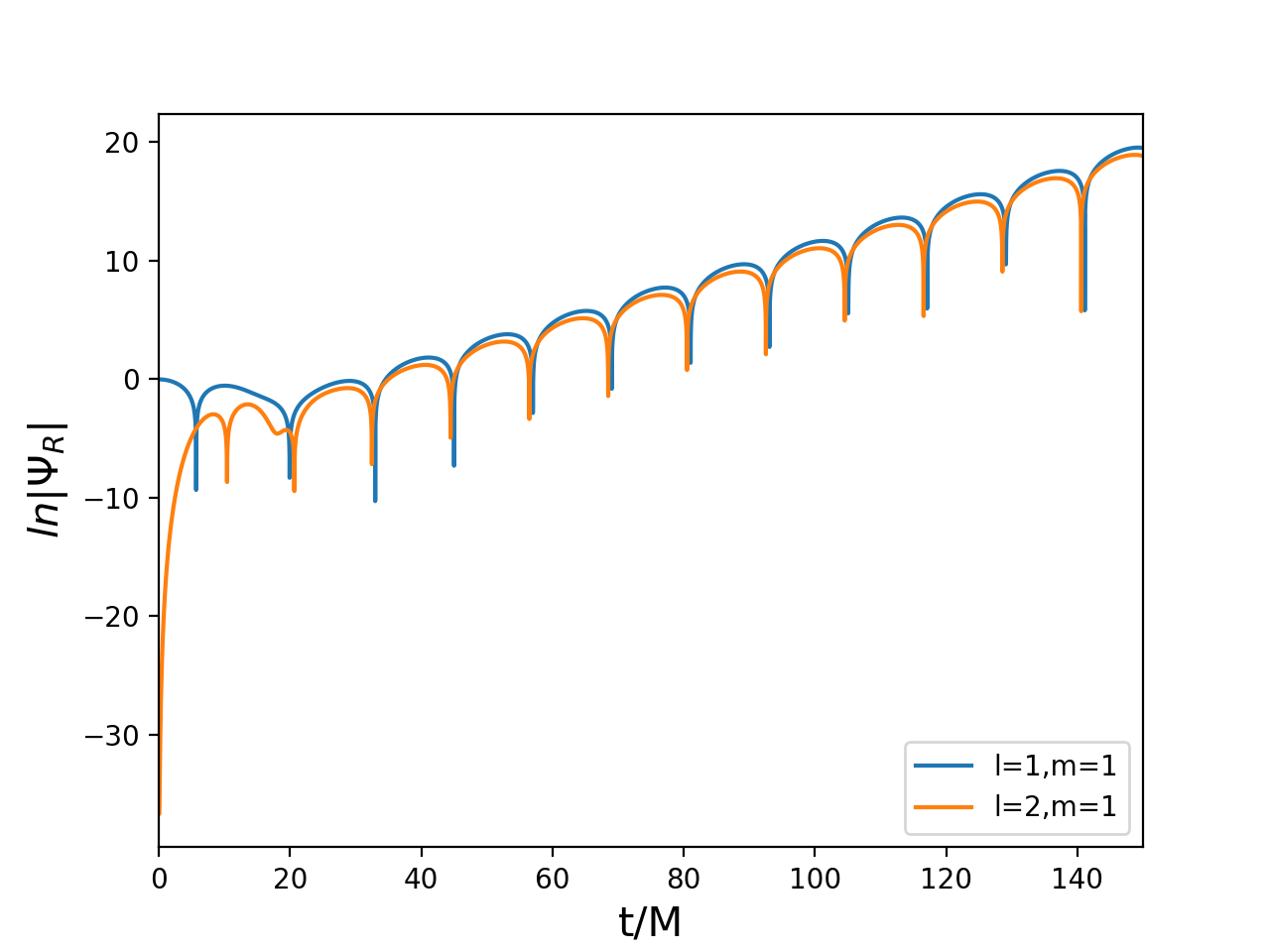}           
	\caption{The time-domain profiles of different modes of scalar perturbation on the Kerr BHs with spin parameter  $a = 0.9$ in the quadratic dCS gravity with coupling constant $\alpha=1$. In the left panel, the initial modes are chosen to be the modes with $m=0$ but $l=0,\;1$, and $2$, respectively, while  the indices $m$ for the initial modes in the right panel are both $m=1$ but $l=1$ and $2$, respectively.  The observing location is at $r\to\infty\ (R=1)$, $\theta=\pi/2$.} \label{fig:m0m1}  
\end{figure}

\begin{figure}        
	\centering 
	\includegraphics[width=0.32\textwidth]  {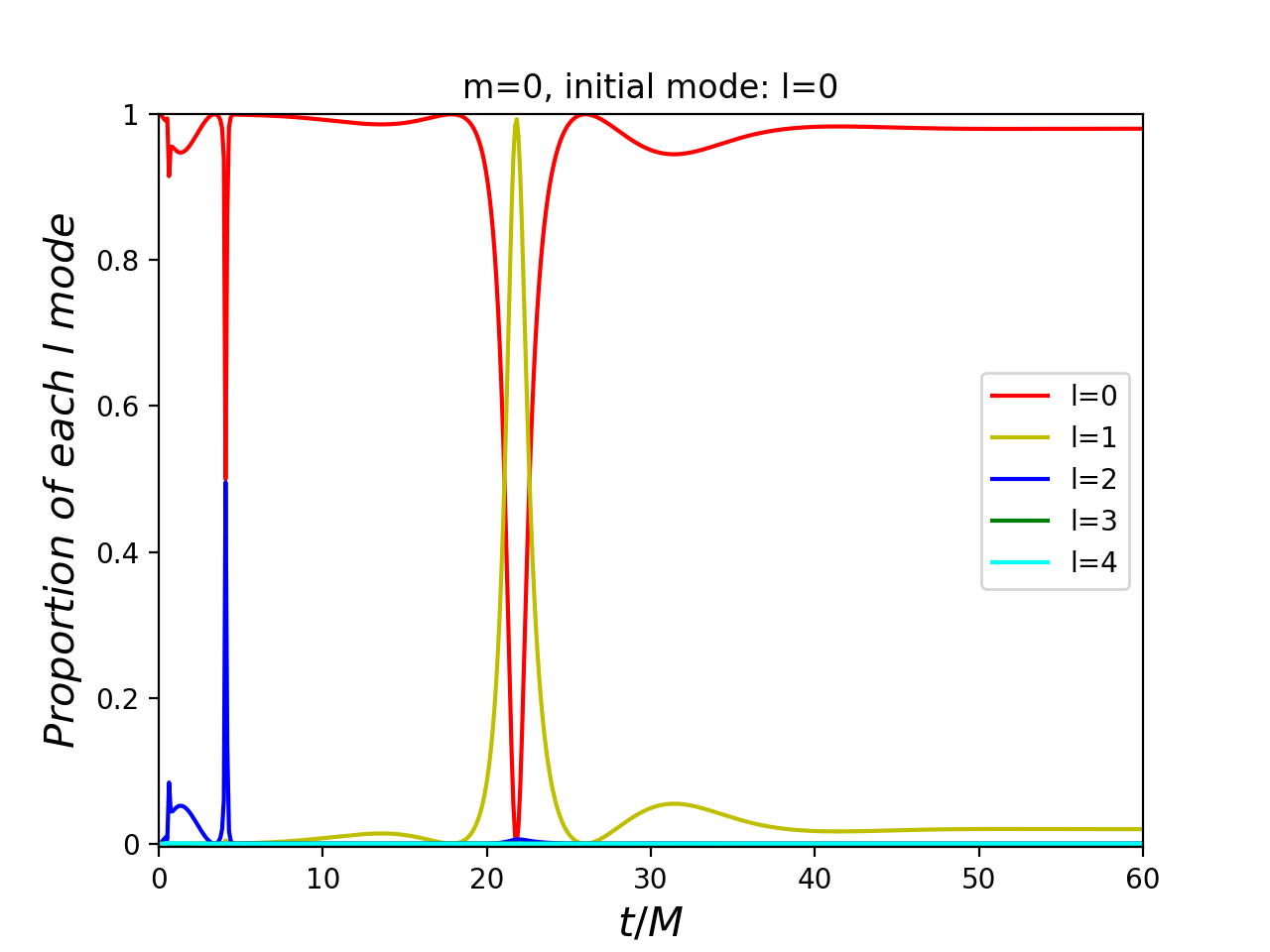}  
	\includegraphics[width=0.32\textwidth]  {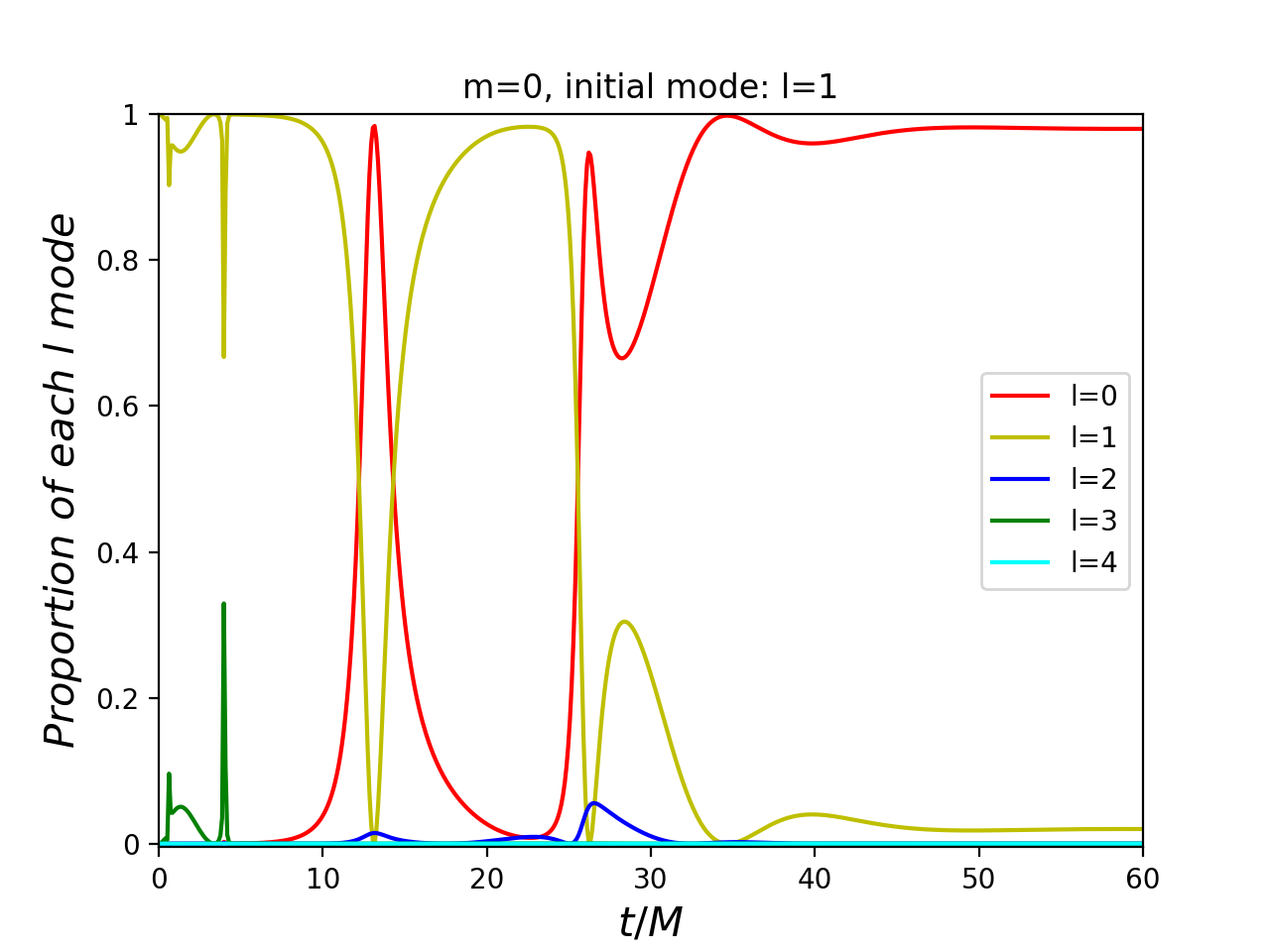}  
	\includegraphics[width=0.32\textwidth]  {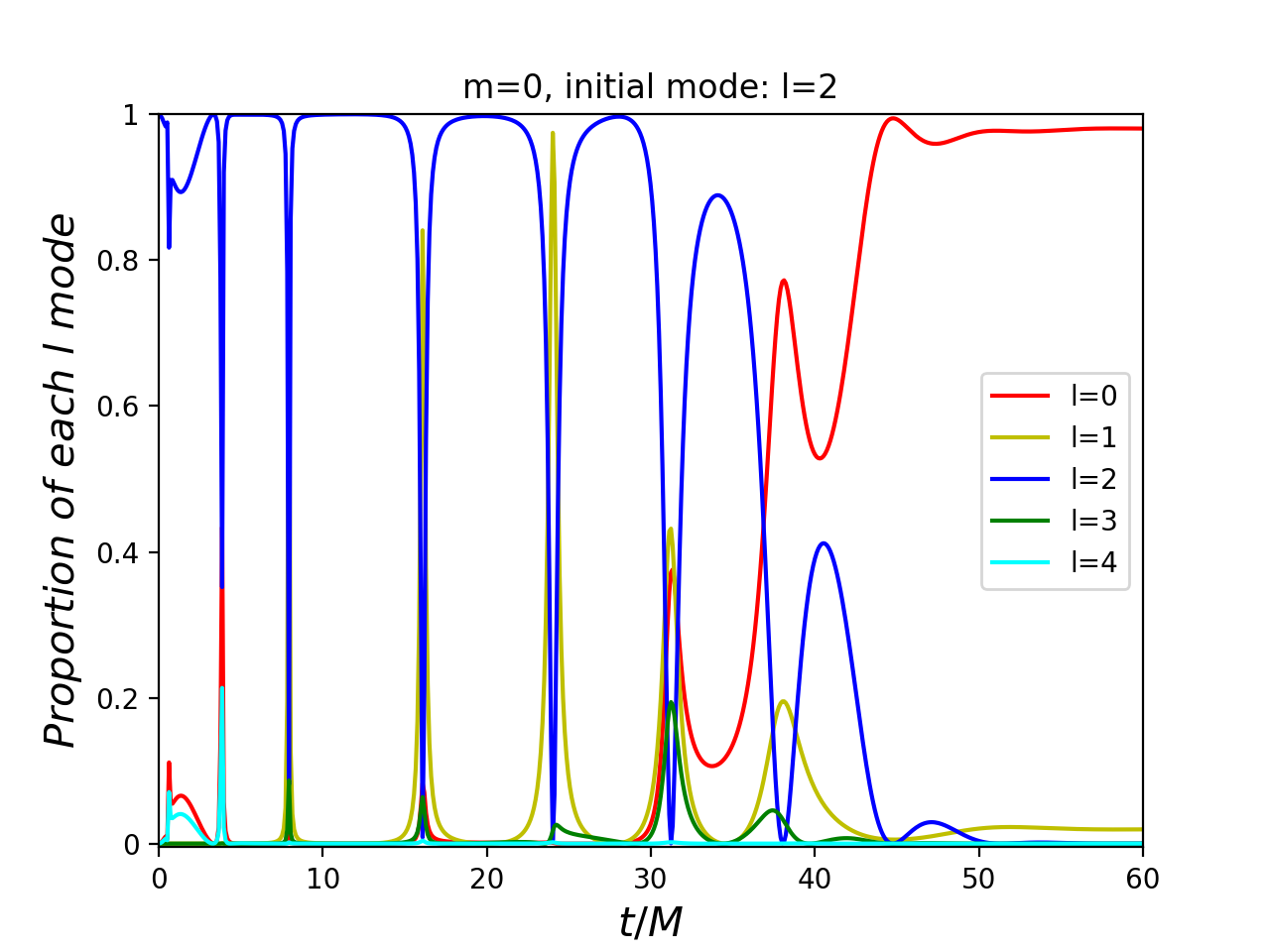}      
	\caption{Several  modes with $m=0$ and different $l$ numbers are activated by the initial $l$ mode during the process of time evolution. From left to right, the initial mode are specified to be the one with $l=0, \;1$ and $ 2$, respectively.  It is found that, at late enough time, the dominant mode is always the $l=m=0$ mode,  whatever the initial mode is chosen. } \label{fig:1a}  
\end{figure}

Before showing our main results, it is helpful to clarify the dependence with angular harmonic numbers $(l,m)$. Note that the PDEs do not explicitly contain $l$, but we should not naively expect that the late-time behavior of the perturbation is theoretically  independent of $l$, although the numerical experiment, as shown in Fig.\ref{fig:m0m1}, seems to indicate that the  modes with the same $m$ but different $l$ numbers have the same  late-time behavior.  In fact, they are indeed different. The reason that  we find the same behavior is that, in the Kerr background,  the different $l$ modes with the same azimuth index $m$ are not independent but coupled with each other. Therefore, although initially there is only one mode with a specified $l$ number,  other $l$ modes with the same index $m$ will be activated during the  process of evolution, and at the late time, the mode with $l=m$ will become dominant. For example, the processes in which several modes with $m=0$ and different $l$ numbers are activated by the given initial  mode with $l=0,\,1$, and $2$ are illustrated, respectively, in the three panels of Fig.\ref{fig:1a}. It is found that, although the initial modes in the three panels have  different $l$ numbers, the late-time dominant mode is always the mode with $l=m=0$.    So  the late-time behaviors shown in  Fig.\ref{fig:m0m1} are just those of the dominant modes with $l=m$.

\subsection{Axisymmetric ($m=0$) modes }

\begin{figure}        
	\centering
	\includegraphics[width=0.48\textwidth]  {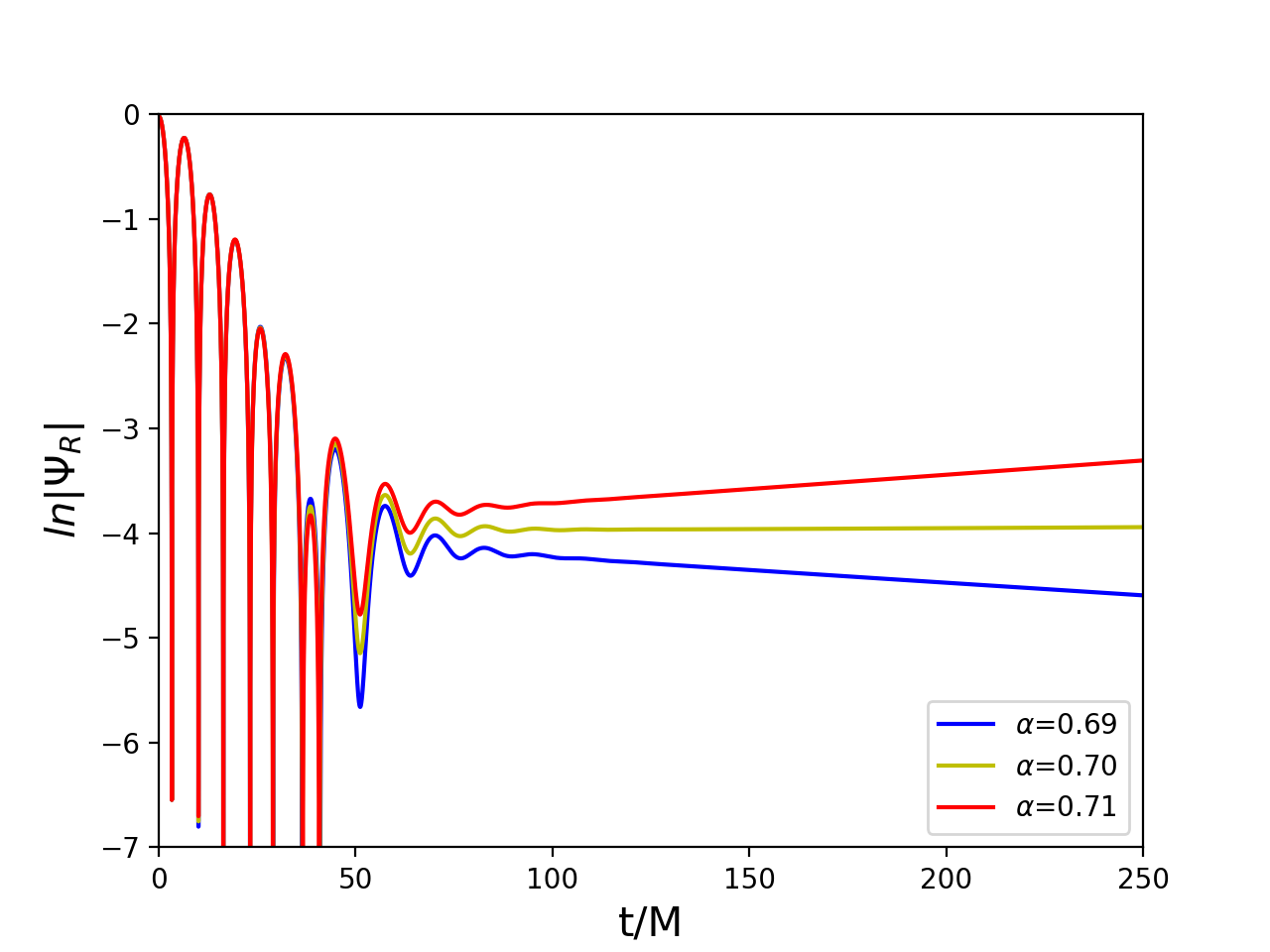}  
	\includegraphics[width=0.48\textwidth]  {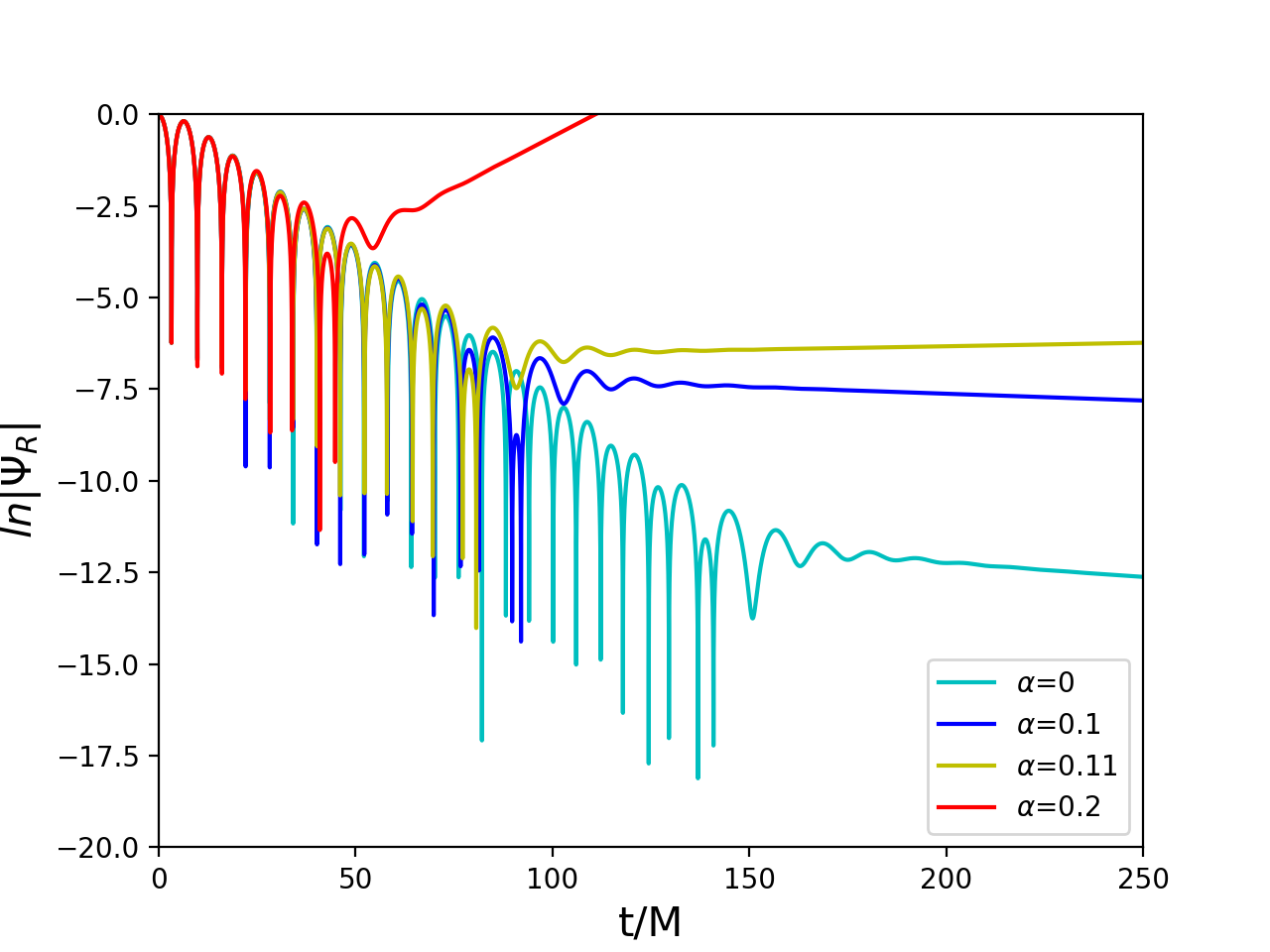}           
	\caption{The time-domain profiles of the dominant $m=0$ mode of scalar perturbation on the Kerr BHs with spin parameter  $a = 0.5$ (left panel) and $0.998$ (right panel) in the quadratic dCS gravity with different values of coupling constant $\alpha$. Here, the initial mode is specified to be $l=2$. The observing location is at $r\to\infty\ (R=1)$, $\theta=\pi/2$.} \label{fig:1}  
\end{figure}

The time-domain profiles of the dominant $m=0$ mode of scalar perturbation around Kerr BHs in quadratic dCS theory with different values of coupling constant $\alpha$ are plotted in Fig.\ref{fig:1}. Here, the initial mode is specified to the one with $l=2$.  As usual in GR, after a  period of damping proper oscillations, dominated by quasinormal modes, a stage of power-law tails  appears at the very late time. Interestingly, for a Kerr BH with given rotation parameter $a$, when the coupling constant $\alpha$ become large enough,  the instability will develop after the damping quasinormal oscillations. The larger $\alpha$ is, the faster the instability grows, and the shorter the period of damping lasts.   It is obvious that the rapidly rotating BH is more susceptible to $\alpha$.  In addition, although the axisymmetric scalar perturbation is coupled with the CS term, our results are still consistent with the fact that in Kerr spacetime, if any unstable mode occurs, it  contains only the imaginary part.

\subsection{$m\neq 0$ modes}

\begin{figure} 
	\centering       
	\includegraphics[width=0.7\textwidth]  {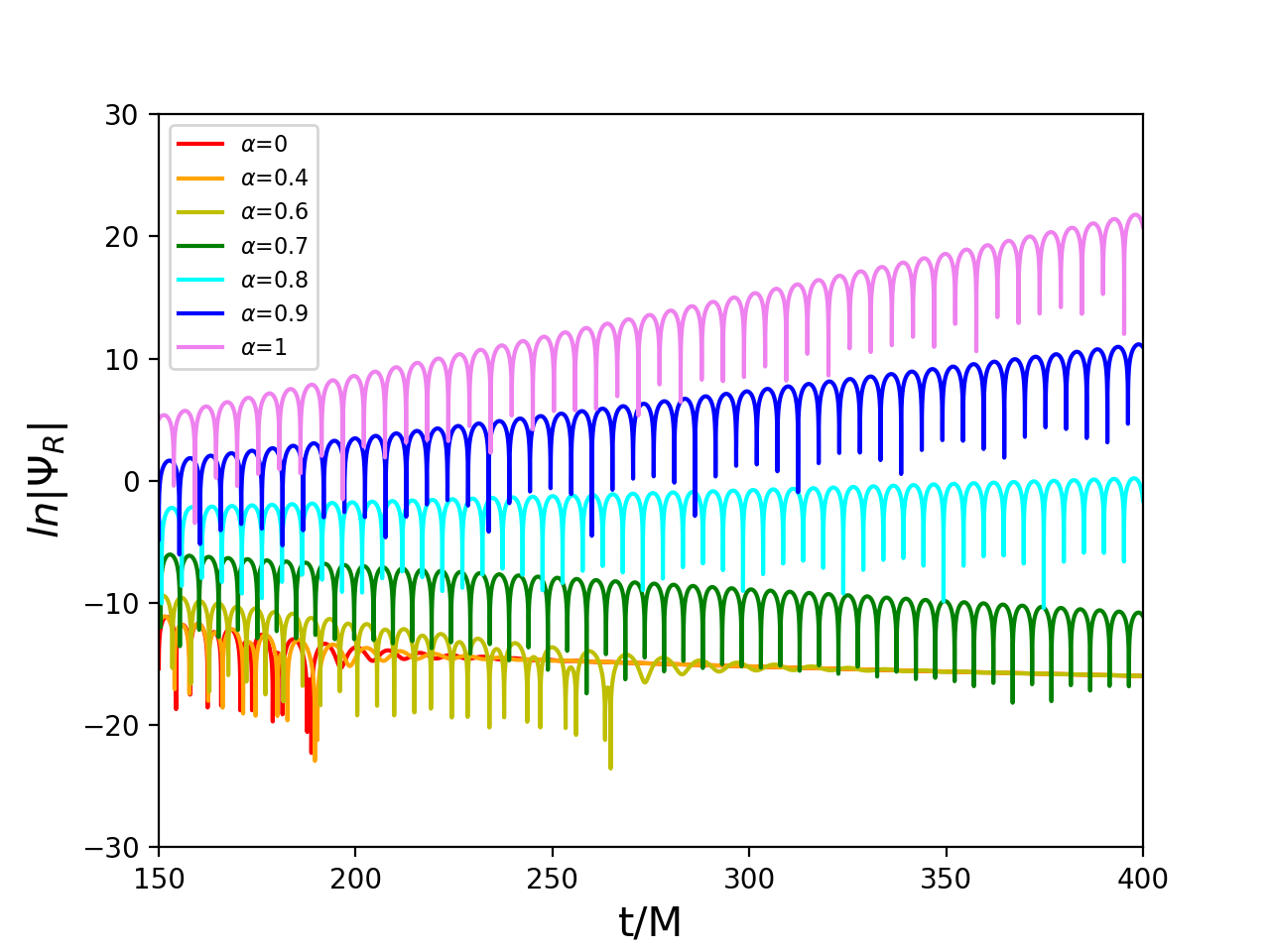}       
	\caption{The late-time evolution of the dominant $m=2$ mode on the background of the Kerr BH with spin parameter $a=0.9$ in theories with different values of coupling constant $\alpha$.}  \label{fig:2}
\end{figure}

When the perturbation is not axisymmetric, there exist $m\neq 0$ modes. In Fig.\ref{fig:2}, we plot the time-domain profiles of the dominant $m=2$ mode on the background of Kerr BH with spin parameter $a=0.9$ in theories with different values of coupling constant $\alpha$. When $\alpha$ is large enough, the mode grows rapidly and instability occurs. Different from the growing axisymmetric ($m=0$) modes, these unstable modes are oscillating and develop immediately after the initial outburst. This is reasonable, actually, in the case of $m=0$, from coefficients of  Eq.\eqref{eq:TE25}, one can easily find that $\psi_R$ and $\psi_I$ are decoupled, there is nothing that can drive the scalar field oscillate, so it just keeps growing  when the unstable mode appears. However, for the cases $m\ne 0$, due to the coupling of $\psi_R$ and $\psi_I$, they interact with each other in the process of time evolution, which allows the scalar field to oscillate as it grows  under unstable patterns.

\begin{figure} 
	\centering       
	\includegraphics[width=0.48\textwidth]  {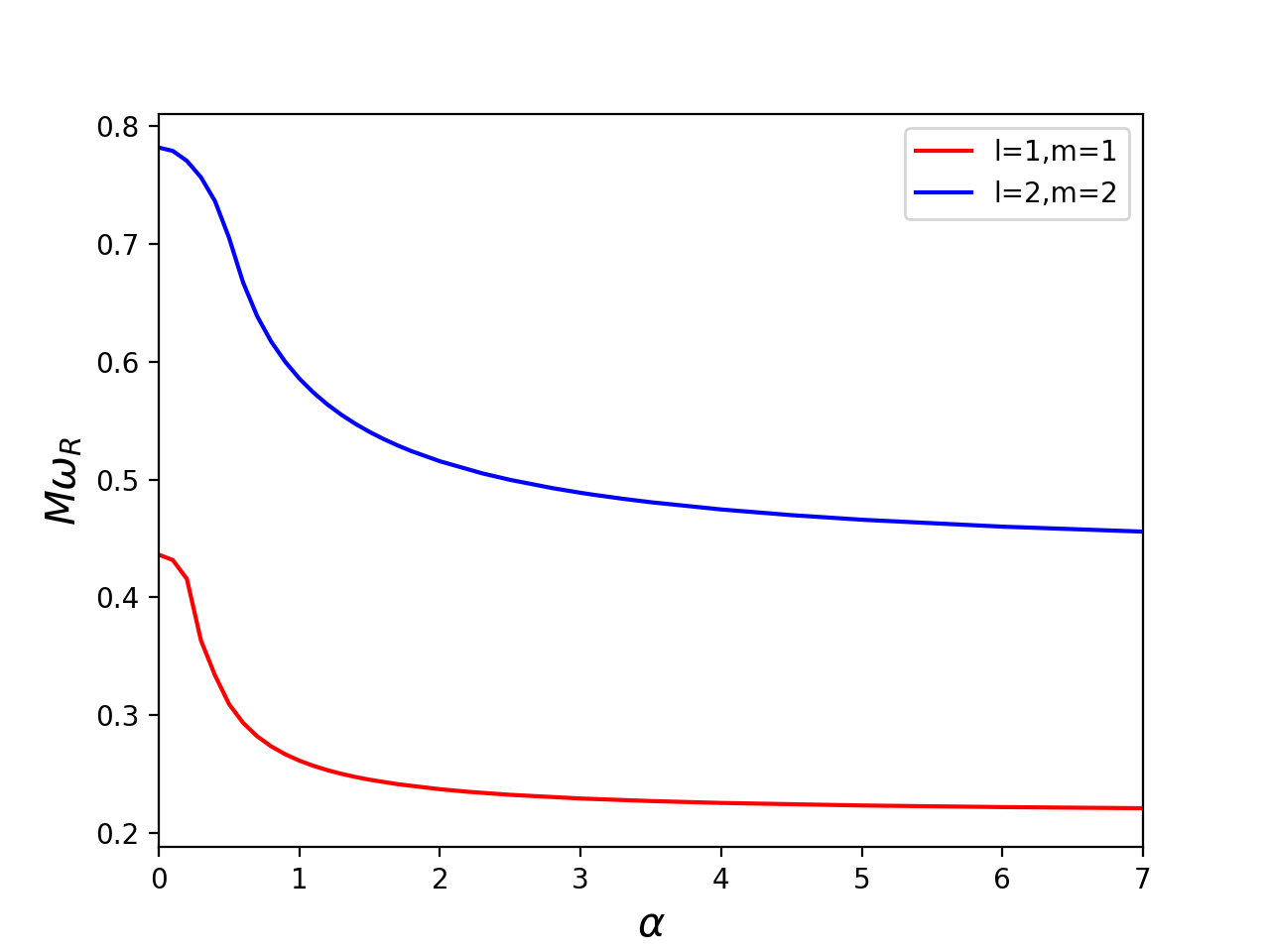}
	\includegraphics[width=0.48\textwidth]  {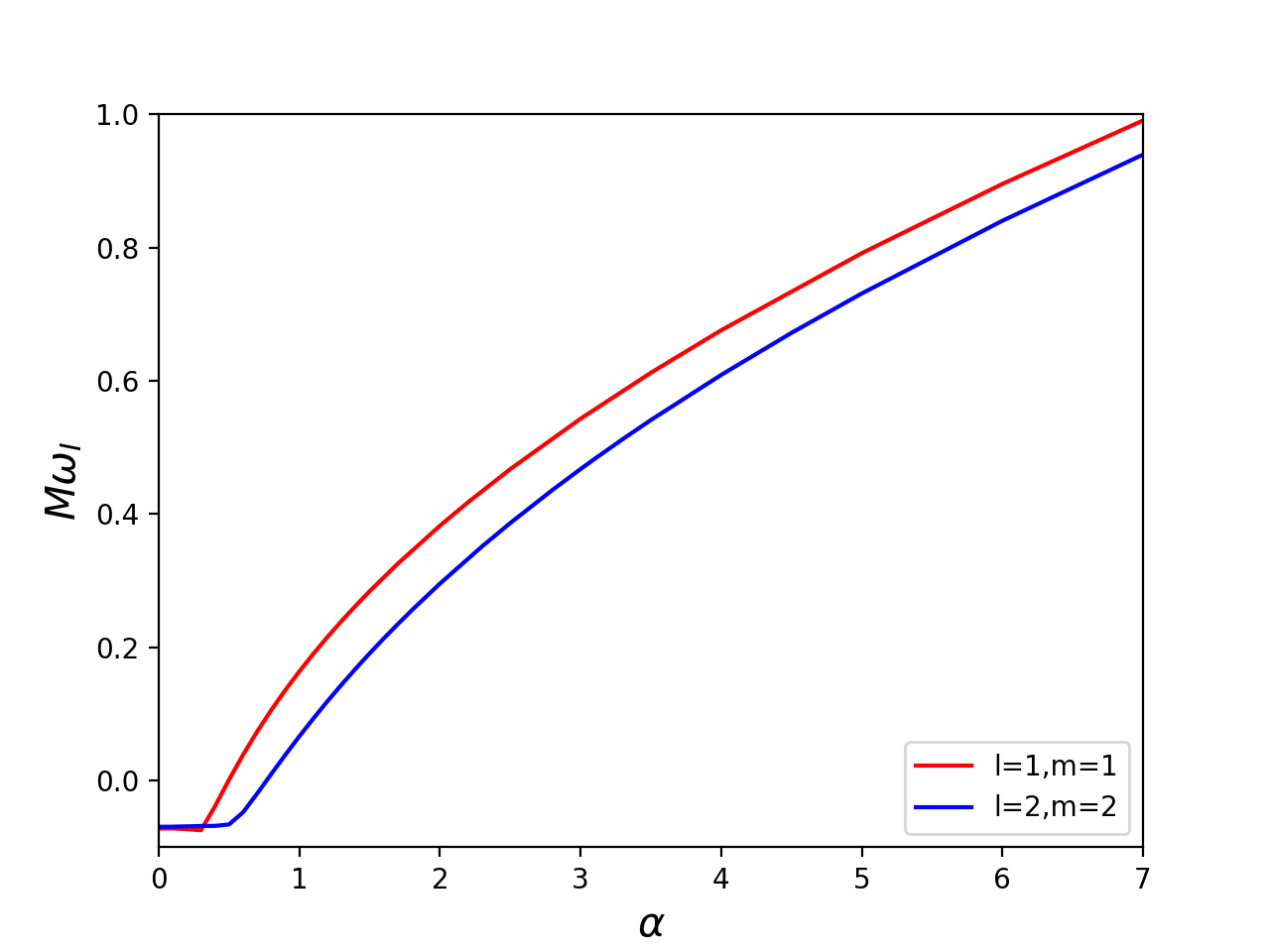}        
	\caption{The real part (left panel) and imaginary part (right panel) of the frequency $\omega$ at the stage of the late-time tail  of dominant $m=1$ ($l=1$) and $m=2$ ($l=2$) modes of the perturbation on the background of a Kerr BH with spin parameter $a=0.9$ in the quadratic dCS gravity with different values of $\alpha$.} \label{fig:RI} 
\end{figure}

In Fig.\ref{fig:RI}, we show the real and imaginary parts of the frequency $\omega$ at the stage of the late-time tail  of  dominant $m=1$ ($l=1$) and $m=2$ ($l=2$) modes of the perturbation on the background of a Kerr BH with spin parameter $a=0.9$, as two functions of the coupling constant $\alpha$ in the left and right panels, respectively. With the increase of the value of $\alpha$, the real part decreases monotonically and tends to a nonzero limit value, which depends on the harmonic azimuthal index $m$; meanwhile, the imaginary part seems to grow monotonically without limit. It is worth noting that the imaginary part can change its sign from a negative value to a positive one for the coupling constant $\alpha \lesssim 1$.  

\subsection{Stable and unstable regions in the parameter space}
\begin{figure}
	\centering       
	\includegraphics[width=0.7\textwidth]  {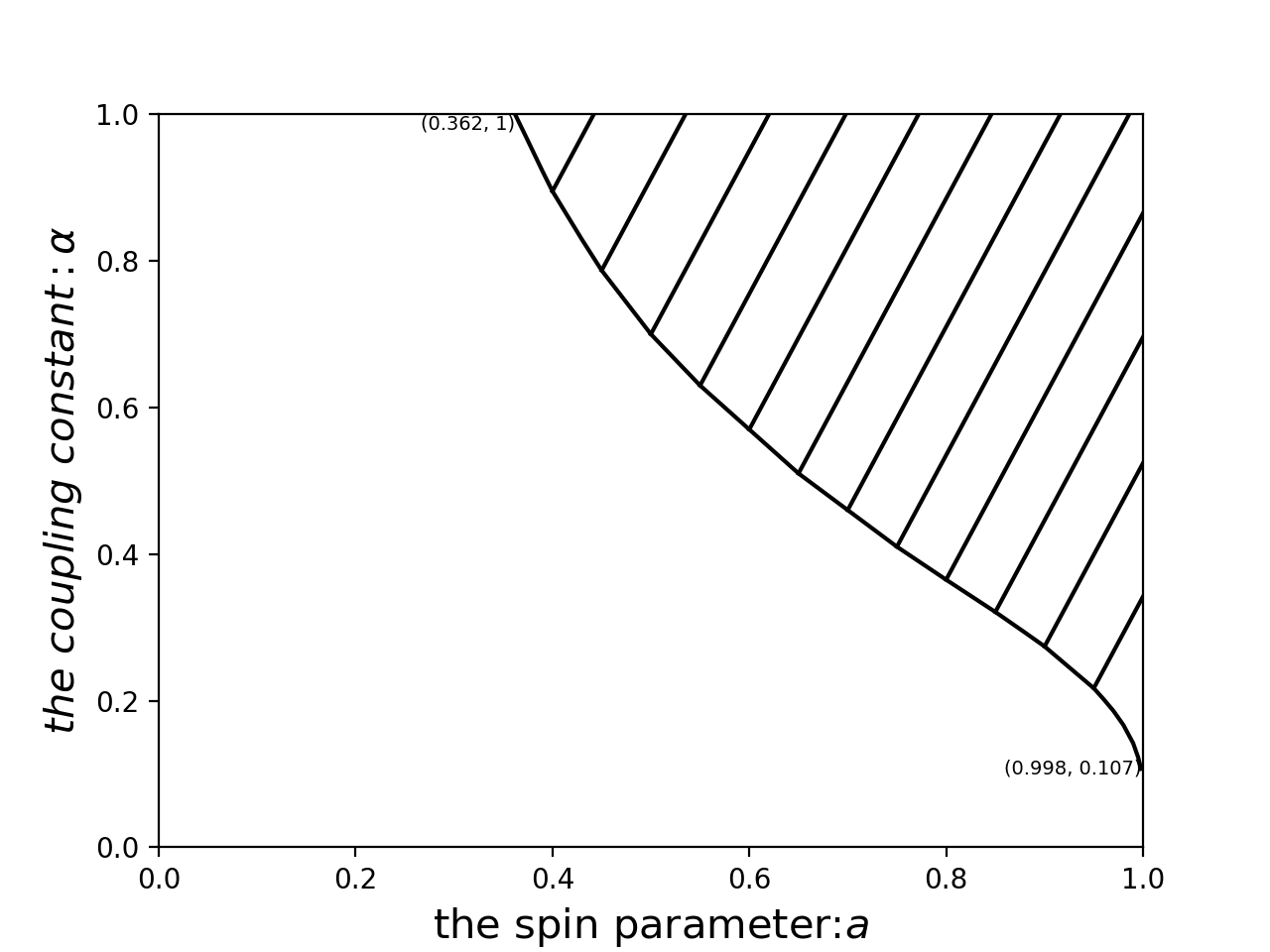}       
	\caption{The parameter space for the dominant $m=0$ mode ($l=0$).  In the region  above this line (shaded),  there exist unstable perturbation modes, and then the BH becomes unstable, while in the region below the line (blank), there is no growing mode and the BH is stable.} \label{fig:3} 
\end{figure}

\begin{figure} 
	\centering       
	\includegraphics[width=0.7\textwidth]  {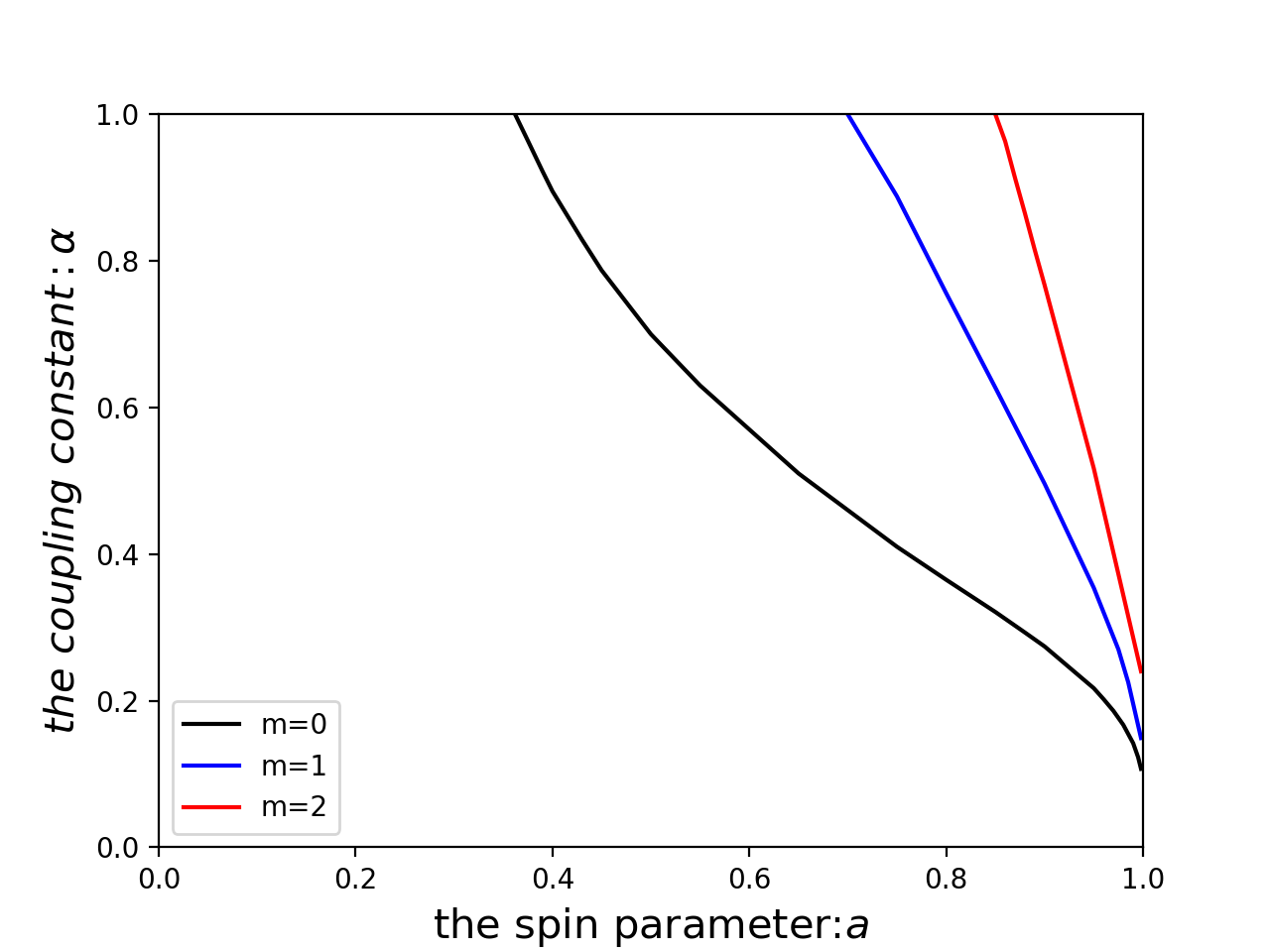}       
	\caption{The boundary between stable and unstable regions in parameter space for different values of harmonic azimuthal index $m$.}
	\label{fig:lines}  
\end{figure}

 From the result above, it is found that the occurrence of instability depends on the value of  spin parameter $a$ and coupling constant $\alpha$. That is to say, this instability occurs only in a certain region of the parameter space. In Fig.\ref{fig:3}, we plot a dividing line in the parameter space spanned by $\alpha$ and $a$ for the $m=0$ mode. For the point in the region  above this line (shaded),  there exist unstable perturbation modes, and then the BH becomes unstable, while for the points in the region below the line (blank), there is no growing mode and the BH is stable.  The boundary between stable and unstable regions depends also on the value of the harmonic azimuthal index $m$, which is illustrated in Fig.\ref{fig:lines}. Obviously, among the dominant $m$ modes, the unstable region for the $m=0$ mode are the largest.  Note that Figs.\ref{fig:3} and \ref{fig:lines} are somewhat similar to Fig. 1 in Ref.\cite{PhysRevD.90.044061}. However, it is worth pointing out that their physical implications are decidedly different.    

\subsection{Reliability tests}
The main result presented above is numerical evidence of an instability. To verify that we are seeing a true instability of the underlying PDEs, rather than a problem with the numerical formulation,  it is necessary to test our code carefully.

For a preliminary check for the validation of our code,  we first consider the degenerated case in which the coupling constant $\alpha$ vanishes. In Table \ref{tb:1}, 
the quasinormal frequencies of the $m=2$ mode with $l=2$ for BHs with different spins are listed.
Our results, listed in the middle column, are obtained in the time domain via the Prony method \cite{PhysRevD.75.124017,Huang2016}. For comparison, the results in the frequency domain presented in Ref.\cite{PhysRevD.76.084001} are also listed in the right column. Clearly, our results are accurate enough. 
\begin{table}[h]
	\centering   
	\caption{Quasinormal frequencies of the $m=2$ mode with $l=2$ for the case in which the coupling constant $\alpha$ vanishes.}
	\begin{tabular}{l|c|c}
		\hline
		&\textbf{Our results}  & \textbf{Results from Ref.\cite{PhysRevD.76.084001}  } \\
		&Re($\omega$)\ \ \ \ \ \ \ \ \ -Im($\omega$)&Re($\omega$)\ \ \ \ \ \ \ \ \ -Im($\omega$)\\
		\hline
		$a=0.1$\ \  &\ 0.499473\ \ \ \ \ 0.096701\ &\ 0.499482\ \ \ \ \ 0.096666 \\
		$a=0.5$\ \  &\ 0.585989\ \ \ \ \ 0.093495\ &\ 0.585990\ \ \ \ \ 0.093494 \\
		$a=0.9$\ \  &\ 0.781638\ \ \ \ \ 0.069287\ &\ 0.781638\ \ \ \ \ 0.069289 \\
		$a=0.995$\ \  &\ 0.949513\ \ \ \ \ 0.023091\ &\ 0.949522\ \ \ \ \ 0.023104 \\
		\hline
	\end{tabular}\label{tb:1}
\end{table}

\begin{figure} 
	\centering       
	\includegraphics[width=0.7\textwidth]{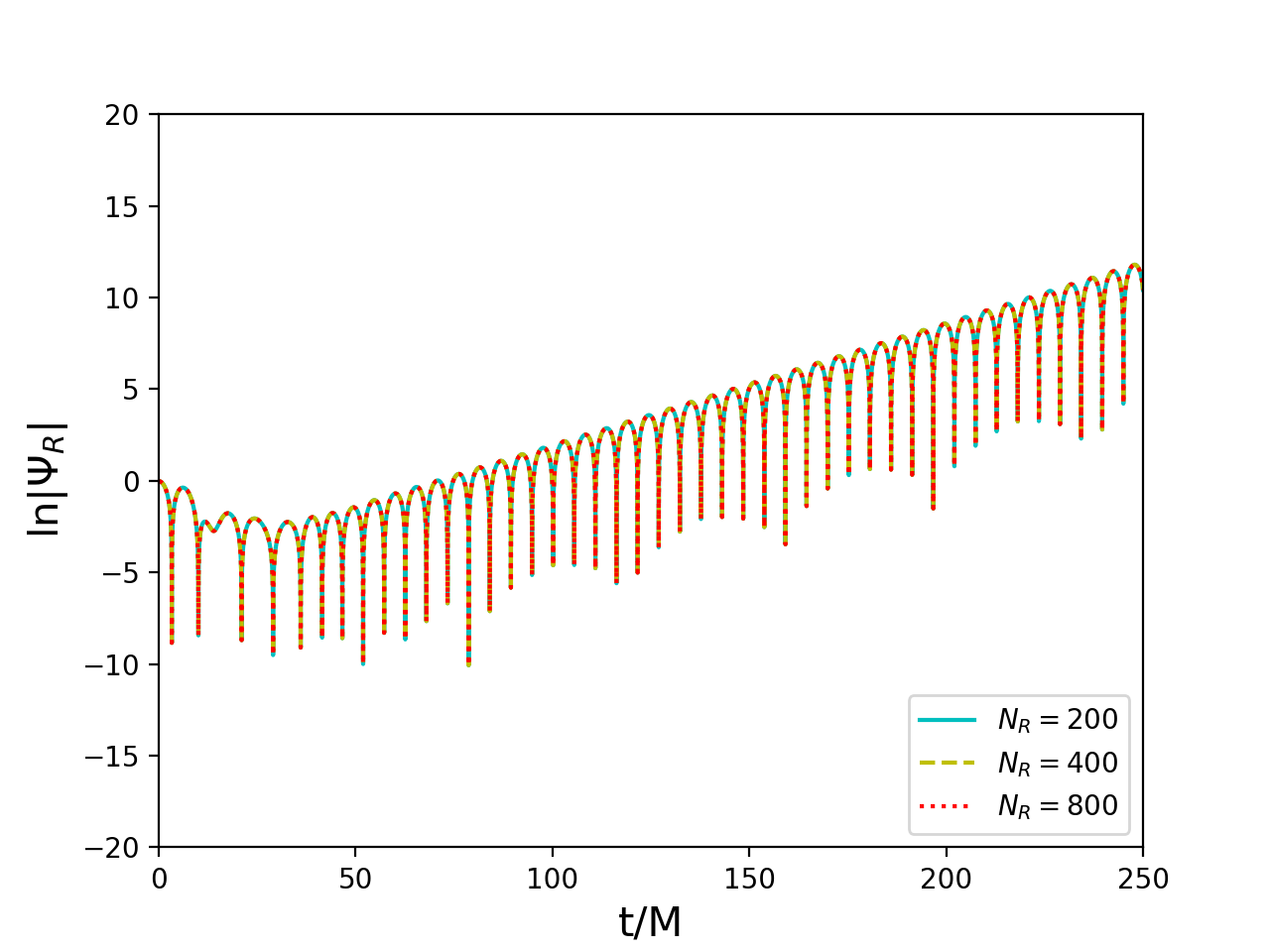}       
	\caption{The time-domain profiles of the dominant $m=2$ mode ($l=2$). Here, we choose three different radial grids $(N_R=200,400,\; \text{and}\; 800)$, and $a=0.9$, $\alpha=1$.}
	\label{fig:3NRs}  
\end{figure}

To further verify the reliability of the code, we need do a convergence test. To this end, we run the code with different radial grids.  The more grids there are, the higher the numerical resolution is. In Fig.\ref{fig:3NRs},  the time-domain profiles of the dominant $m=2$ mode ($l=2$) are  plotted with the numbers of radial grids $N_R=200, 400$, and $ 800$, respectively.  From  Fig.\ref{fig:3NRs}, we find that  the late-time evolution is almost the same in the three cases, so it can be concluded that the values of frequency converge with increasing numerical resolution.  

\begin{figure} 
	\centering       
	\includegraphics[width=0.7\textwidth]{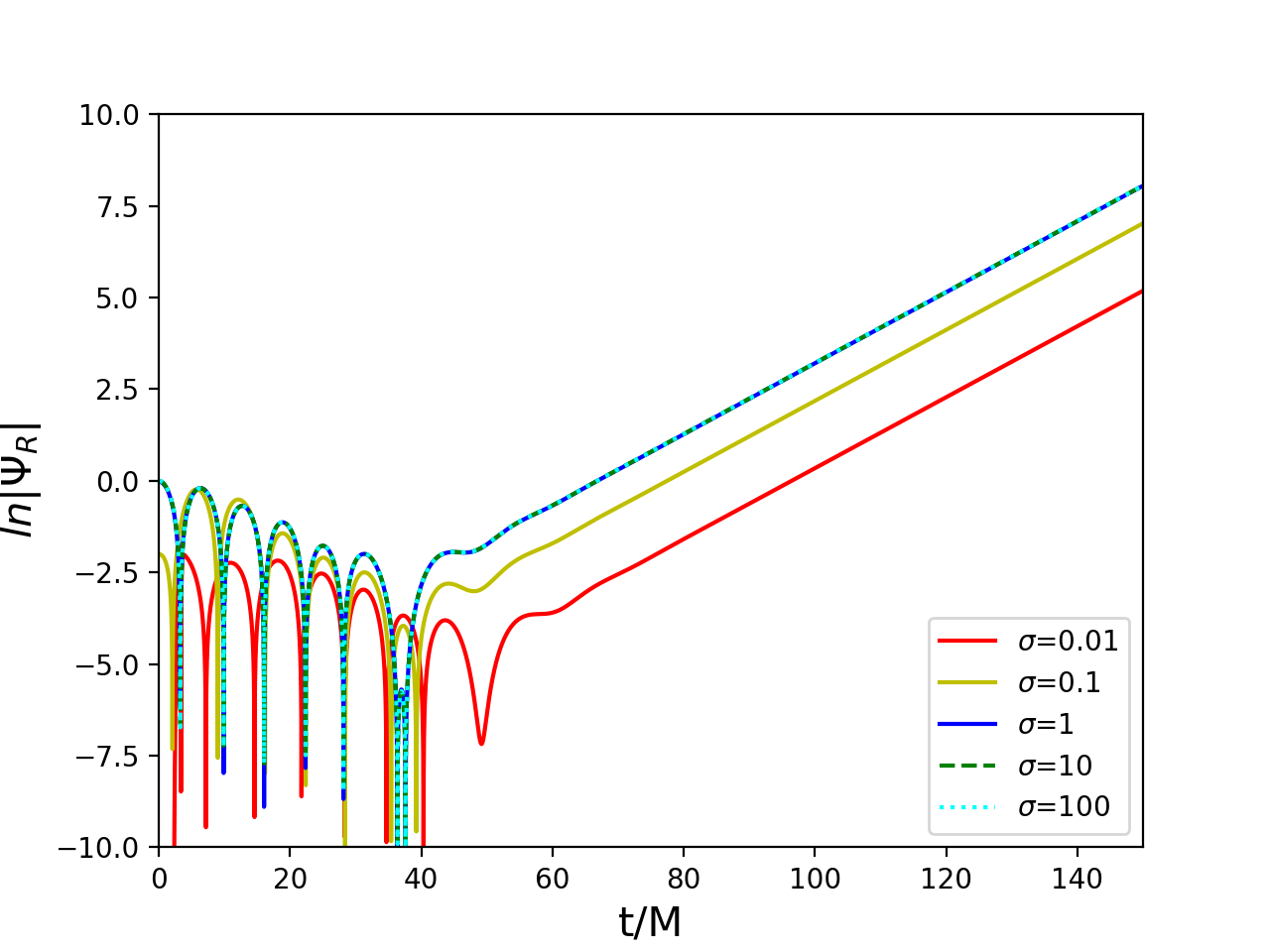}       
	\caption{The time-domain profiles of the $m=0$ mode for initial wave packets with  parameter $\sigma=0.01,0.1,1,10$, and $100$. Here, we set $a=0.9$, $\alpha=0.5$, and the initial mode $l=2$. Clearly, the rate of blowup does not change for a large range of $\sigma$.}
	\label{fig:ICsigma}  
\end{figure}

Next, we want to check whether the instability depends on the initial conditions.  For this purpose, we run the code by choosing different values of parameter $\sigma$ of the initial wave packet \eqref{eq:IC}. As is shown in Fig.\ref{fig:ICsigma}, for a wide range of $\sigma$, the rate of blowup does not change, which indicates that the instability always
presents at the same rate regardless of initial conditions.

\begin{figure} 
	\centering       
	\includegraphics[width=0.7\textwidth]{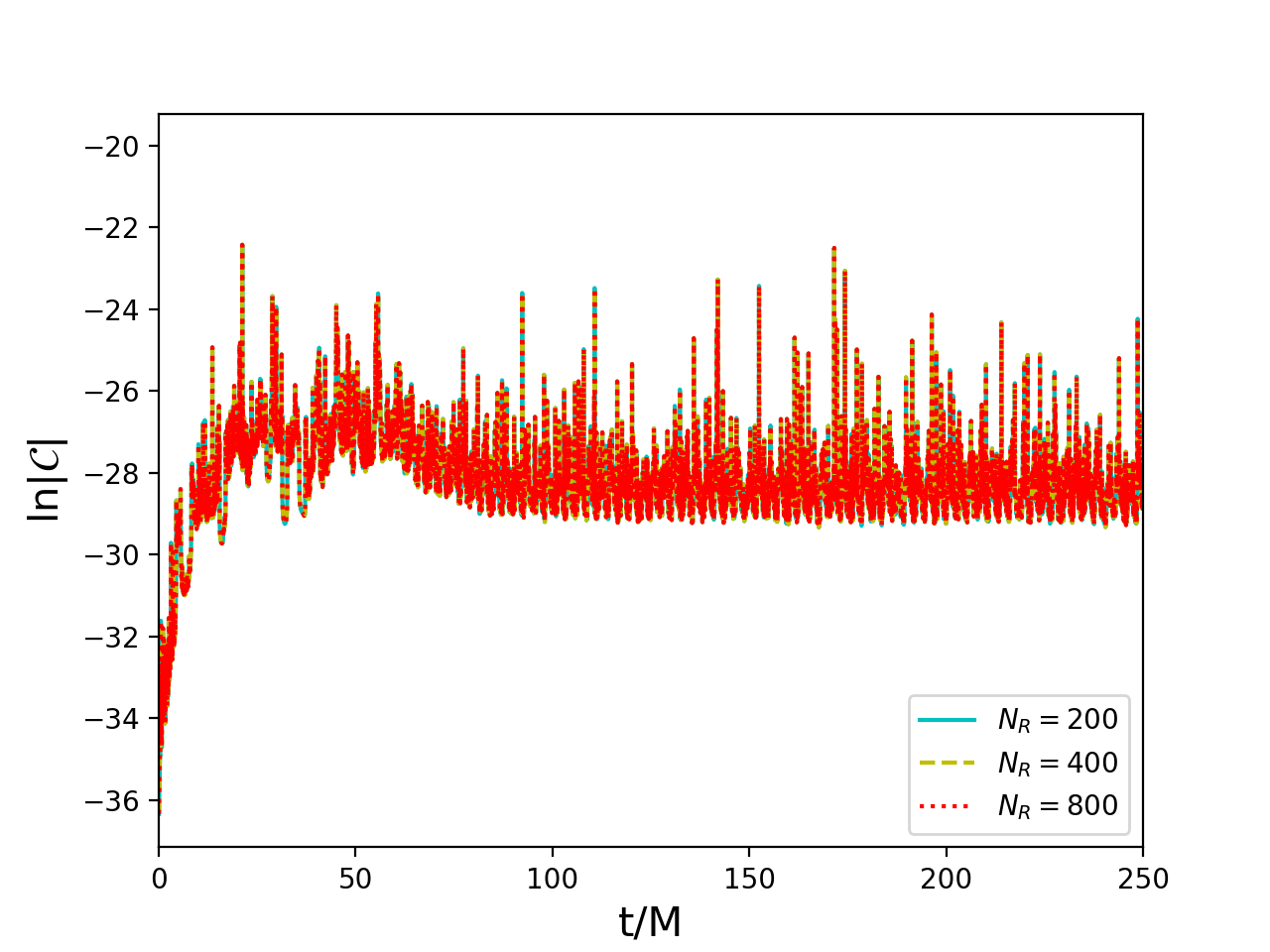}       
	\caption{The absolute value of relative difference between the auxiliary field $\Gamma$ and $\partial_{R}\psi$. The parameters are selected to be the same as those in Fig.\ref{fig:3NRs}.}
	\label{fig:constraints}  
\end{figure}

Last but not least, adding an auxiliary field usually means that there is a constraint in the enlarged configuration space. To illustrate conveniently that the constraint is satisfied throughout the evolution, we take the $\partial_R \psi$ as another auxiliary field $\Gamma$, which evolves as
	\begin{equation}
	\partial_T \Gamma=\partial_{R}\Pi-b\partial_{R}\Gamma-(\partial_{R}b)\Gamma,
	\end{equation}
and define  the relative difference between  $\Gamma$ and $\partial_{R}\psi$ as a measure of the constraint
\begin{equation}
\mathcal{C}=\frac{\Gamma-\partial_{R}\psi}{\Gamma}.
\end{equation}
In Fig.\ref{fig:constraints},  the absolute value of $\mathcal{C}$ as a function of time $t$ is plotted. Initially, there is no difference between $\Gamma$ and $\partial_{R}\psi$, as is specified. Because of the numerical errors, the difference is made during the evolution. Obviously, the difference is very small throughout the evolution, which provides evidence that the constraints are always satisfied. 
	
All the results of numerical tests indicate that our code is reliable and the instability we find is not due to the numerical method but a true instability of the underlying physics. 

\section{Instability and Validity of the EFT}
\label{EFT}

As shown by many authors (see, e.g., Refs\cite{Delsate2014,PhysRevD.86.124031,PhysRevD.90.044061}), dCS gravity should be considered as an EFT of a more fundamental theory. As such, it possess a cutoff scale beyond which its action should be modified through the inclusion
of higher-order curvature terms.  The same goes for the quadratic dCS gravity. The cutoff length scale $\lambda_c$ for the theory, which can be
determined by estimating the order of magnitude of loop
corrections to the coupling term in Eq.\eqref{eq:action}, reads
\begin{equation}
\lambda_c= L_{\text{pl}}^{1/2}\tilde{\alpha}^{1/4},
\end{equation}
where $L_{\text{pl}}$ is the  Planck length scale. If we take $\lambda_c \sim 10 \mu\mathrm{m}$ to meet the bound of tabletop experiments \cite{PhysRevLett.98.021101}, it is found that $\alpha^{1/2} \lesssim 10^{22}\, \mathrm{km} $, which is much less restrictive than that for the theory with linear coupling \cite{Yagi:2012ya,PhysRevD.86.124031}.  Note that this cutoff implies that not all $k$ numbers of the field excitation are within the regime of validity of this EFT; we must have $k \lesssim 1/\lambda_c$. 

Dyda \emph{et al}. \cite{PhysRevD.86.124031} have shown that  dCS theory with linear coupling suffers a ghost instability for the modes with a sufficiently-high wave number. However, this instability is at or beyond the cutoff scale of the EFT. Similarly, Stein \cite{PhysRevD.90.044061} showed that the coupling parameter in the linearly coupled case is bounded from above depending on the spin of the BH, in order for the correction to Kerr BH to be under perturbative control.  Therefore, an interesting question arises: Is the scalar instability found in this work within the regime of validity of the EFT?

%As is mentioned previously,  the quadratic dCS gravity we have studied should  be treated as an EFT just like  dCS theory with linear coupling. 

To answer this question, we take the decoupling limit, assuming that the corrections due to the interaction term are small \cite{PhysRevD.90.044061}. 
That is, we take $\tilde{\alpha} \to \varepsilon \tilde{\alpha}$
and expand both the  field $\Phi$ and  metric $g_{\mu\nu}$ in powers of $\varepsilon$, 
\begin{align}
\Phi&=\Phi_0+\varepsilon\Phi_1+\varepsilon^2\Phi_2+\cdots\\
g_{\mu\nu}&=g_{\mu\nu}^{\text{GR}}+\varepsilon h_{\mu\nu}^{(1)}+\varepsilon^2h_{\mu\nu}^{(2)}+\cdots
\end{align}
where $\Phi_0$ is a constant and $g_{\mu\nu}^{\text{GR}}$ denotes a GR solution for the metric. This can be done because in the limit $\varepsilon\to 0$ the theory goes back to GR.  Since the leading terms of both $\tilde{\alpha}C^{\mu\nu}$  and $T_{(\Phi)}^{\mu\nu}$ are of order $\varepsilon^2$,  we obtain from Eq.\eqref{eq:Einstein} that
\begin{align}
G_{\mu\nu}[h_{\mu\nu}^{(1)}]=0
\end{align}
at the order of $\varepsilon$, where $G_{\mu\nu}$ is the Einstein-Hilbert operator of the background acting on the metric perturbation. This is just the metric perturbation equation in a background represented by $g_{\mu\nu}^{\text{GR}}$.  It is known that the perturbation $h_{\mu\nu}^{(1)}$ will decay quickly and then vanish in the background of Kerr or Schwarzschild spacetime. So, the leading-order metric deformation away from GR enters at $\varepsilon^2$, which satisfies the equation
\begin{equation}
G_{\mu\nu}[h_{\mu\nu}^{(2)}]+\frac{8\tilde{\alpha}\Phi_0}{\kappa}C_{\mu\nu}[\Phi_1]=\frac{1}{2\kappa}T_{\mu\nu}[\Phi_1].
\end{equation}
For our purposes, we only need to take the trace of the above equation. In the Lorentz gauge, $\nabla^{\nu} h_{\mu\nu}^{(2)}=\frac{1}{2}\nabla_{\mu}h^{(2)}$, we have
\begin{equation}\label{eq:h2}
\kappa\, \Box h^{(2)}=-\left(\nabla^{\mu}\Phi_1\right)\left(\nabla_{\mu}\Phi_1\right),
\end{equation}
where $h^{(2)}\equiv g_{\mathrm{GR}}^{\mu\nu}h_{\mu\nu}^{(2)}$.
Meanwhile,  the leading deformation to the scalar field $\Phi_1$ is determined by
\begin{equation}\label{eq:phi1}
\Box \,\Phi_1=-2\tilde{\alpha}\Phi_0\;^{\ast}RR^{(0)},
\end{equation}
where superscript $(0)$ means that $^{\ast}R R$ is taken to the zeroth order.
The criterion for the perturbation to be under control can be  set by  
\begin{equation}\label{eq:criterion1}
|h^{(2)}|\lesssim 1.
\end{equation}

Notice that Eqs.\eqref{eq:h2} and \eqref{eq:phi1} are, respectively, the same as Eqs. (11) and (9) in Ref.\cite{PhysRevD.90.044061}, except for the different coefficients of the CS invariant term. Therefore, following the same scaling procedure as in Ref.\cite{PhysRevD.90.044061}, we find that for the background of a constant field $\Phi_0$ together with a Kerr BH with mass $M$ and spin $a$ the coupling constant should satisfy
\begin{equation}\label{eq:constraint}
\frac{\tilde{\alpha}}{M^2}\lesssim \frac{\sqrt{\kappa}}{2\Phi_0 \chi(a)}
\end{equation}
in order to keep the corrections to background  under control.  Here, $\chi(a)$ is the square root of the maximum value of  $\tilde{h}$ determined by Eqs. (22) and (24) in Ref.\cite{PhysRevD.90.044061}, which  is a monotonic increasing dimensionless function  of $a$. Moreover, $\chi(a)\sim 1$ for $a\sim M$, and $\chi(a)$ vanishes in the $a\to 0$ limit. The most remarkable feature of this constraint is that it depends on the background value of the scalar field $\Phi_0$.  The smaller the value of $\Phi_0$ is, the weaker the constraint on  the coupling constant $\tilde{\alpha}$ are.  
So, it seems that  the scalar instability we find is always within the regime of EFT validity, as long as $\Phi_0$ is small enough.

%Roughly speaking, the above constraints are obtained  by ensuring that the coupling term  $\tilde{\alpha} \Phi^2 \,^{\ast}RR$ is small compared to the Einstein-Hilbert term, $\kappa R$. %this is just a necessary condition, there are more stringent conditions.  
However, for the EFT to be valid,  it should be also ensured that the scalar field is  in the weak-coupling regime.  In other words, the quadratic coupling term  should be more important than potential higher-order operators that have been truncated from the action. Before we compare the possible higher-order operators with the coupling term, let us first estimate the magnitude of the coupling term.

To be convenient and dimensionally correct, let us introduce a mass scale $\mathcal{M}$ and  
further set the reduced Planck constant $\hbar=1$. Then, the quadratic coupling term, which is a dimension-6 operator, can be written as  $c_1 \dfrac{\Phi^2}{  \mathcal{M}^2} \,^{\ast}RR$, where $c_1$ is a dimensionless coupling constant of order unity. Estimating the curvature $R\sim \dfrac{M}{r^3}$ at a distance $r$ from the
black hole with mass $M$ and assigning a $k$ number to the scalar-field
excitation,  we have $$\dfrac{\Phi^2}{  \mathcal{M}^2} \,^{\ast}RR \sim  k \dfrac{\Phi^2}{ \mathcal{ M}^2}  \dfrac{M^2}{r^5} $$ modulo a surface term. Here, one derivative has been moved onto $\Phi^2$ because the CS invariant $\;^{\ast}RR$ is locally divergent.

%To the next-order operators compared with the quadratical coupling term,  One example is $\dfrac{\Phi^7}{\mathcal{M}^3}$. In order to avoid strong coupling,

In principle, we should compare all higher-order operators with the interaction term, and find the most stringent constraint on the regime of validity of this EFT. However, for our purposes, it is enough to consider the dimension-7 operators such as
\begin{equation}\label{eq:dim7operators}
\dfrac{\Phi^7}{\mathcal{M}^3}, \quad \dfrac{\Phi^3} { \mathcal{M}^3} (\nabla_a \Phi) (\nabla^a \Phi) \quad \mathrm{and} \quad\dfrac{\Phi^3} { \mathcal{M}^3}\,^{\ast}RR.
\end{equation}
To avoid strong coupling, we must keep them to be much smaller than $k\dfrac{\Phi^2}{ \mathcal{ M}^2}  \dfrac{M^2}{r^5}$.
Therefore, we obtain, respectively, that
$$\frac{\Phi^5}{\mathcal{ M}} \ll k\frac{M^2}{r^5},\quad  k \frac{\Phi^3}{\mathcal{ M}} \ll\frac{M^2}{r^5}\quad \mathrm{and} \quad \Phi \ll \mathcal{ M},$$ which imply that 
$$\frac{k}{\mathcal{ M}}\ll \left(\frac{M^2}{r^5 \mathcal{ M}^3}\right)^{1/4}\;\mathrm{and}\quad\frac{\Phi}{\mathcal{ M}}\ll \left(\frac{M^2}{r^5 \mathcal{ M}^3}\right)^{1/4}\quad \mathrm{if}\quad \dfrac{M^2}{r^5 \mathcal{ M}^3}\lesssim 1. $$
% Clearly, $\Phi$ should not have too large of a $k$-number,and it should not be displaced too far from the origin in field space.  
Clearly, once the scalar field $\Phi$ has a large $k$ number or is displaced too far from the origin in field space, the EFT treatment becomes invalid. It is worth pointing out that in the linearly coupled theory no constraint is obtained on how far the scalar field can be displaced from the origin in field space because in that case the theory is shift symmetric and so  only shift-symmetric higher-order operators are considered.  However, in the quadratically coupled case, since there is no shift symmetry, many possible operators as listed in Eq.\eqref{eq:dim7operators} have to be considered. This is why we get constraints on how far the scalar field can be displaced. Note that these conditions depend on the distance to the black hole. It is expected that, far away from the black hole, if $\Phi$ grows exponentially due to the instability, the conditions for the EFT treatment will be inevitably broken. Therefore, at least on the linear level, the scalar instability we find will go beyond the regime of validity of the EFT.
 
\section{conclusions}
\label{sec:conclusion}
In this paper, the behavior of time evolution for scalar perturbations on the background of Kerr BHs in the dCS gravity with a quadratic coupling term has been numerically investigated in detail. It was found that under scalar perturbations the Kerr BH becomes indeed unstable at the linear level in some specific region of parameter space, which depends on the value of harmonic azimuthal index $m$ and may break the condition for the theory to be considered an EFT.
 
The instability that we have found can be understood straightforwardly from the fact that, differently from that in usually linearly coupled dCS theory, $\tilde{\alpha}\,^{\ast}RR$ in this quadratically coupled case plays a role  like a position-dependent mass squared in the scalar-field equation. However, the quantity $\,^{\ast}RR$ is not positive definite; whatever the rotation of the black hole, above one of the two poles, $\,^{\ast}RR$ will be positive, and above the other pole, $\,^{\ast}RR$ will be negative.  Therefore high-frequency modes that essentially see $\,^{\ast}RR$ as constant over a few wavelengths will
experience a tachyon instability.

In recent years, there have been lots of studies on spontaneous scalarization, based on the unstable mode of the BH under kinds of perturbations. Most of them were done in the Schwarzschild background. Our work is also connected with them, although it is not sure whether the regime of this instability is spontaneous scalarization, we can almost exclude the possibility of superradiance, since we have found that the instability can occur even in the case of $m=0$.

Finally, what will be the final state of the instability when nonlinear effects are taken into account? It is expected that the Kerr BH will undergo spontaneous scalarization, thus forming a non-Kerr BH with scalar hair. Perhaps a fully nonlinear analysis is required to confirm this possibility. Obviously, this deserves new work in the future. 

\begin{acknowledgments}
We thank the anonymous reviewer for their helpful comments and suggestions. This work is supported in part by the Science and Technology Commission of Shanghai Municipality
under Grant No. 12ZR1421700 and the Program of Shanghai Normal University KF201813.
\end{acknowledgments}

\bibliographystyle{apsrev4-1}
\bibliography{Refs} 
\end{document}